\documentclass[final,3p,times,twocolumn]{elsarticle}

\usepackage{amssymb}
\usepackage{amsmath} 

\usepackage{soul}        
\usepackage{subcaption}  
\usepackage{comment}     
\usepackage{acronym}     
\usepackage{booktabs}    
\usepackage{multirow}    
\usepackage{mathtools}
\usepackage{hyperref}    
\usepackage{xspace}      
\usepackage{algorithm2e} 
\usepackage[inline]{enumitem} 
\graphicspath{ {} }

\usepackage{xcolor} 
\acrodef{acopf}[ACOPF]{alternating current optimal power flow}
\acrodef{cpu}[CPU]{central processing unit}
\acrodefplural{cpu}[CPUs]{central processing units}
\acrodef{gpu}[GPU]{graphical processing unit}
\acrodefplural{gpu}[GPUs]{graphical processing units}
\acrodef{kkt}[KKT]{Karush-Kuhn-Tucker}
\acrodef{simd}[SIMD]{single-instruction multiple-data}
\acrodef{gcd}[GCD]{graphics complex die}
\acrodefplural{gcd}[GCDs]{graphics complex dies}
\acrodef{olcf}[OLCF]{Oak Ridge Leadership Computing Facility}

\newcommand{\nvidia}{NVIDIA\xspace}
\newcommand{\hiop}{HiOp\xspace}
\newcommand{\exago}{ExaGO\textsuperscript{TM}\xspace}

\newcommand{\gko}{\textsc{Ginkgo}\xspace}

\journal{IJEPES}

\begin{document}

\begin{frontmatter}



\title{GPU-Resident Sparse Direct Linear Solvers for Alternating Current Optimal Power Flow Analysis}



\author[inst1]{Kasia \'{S}wirydowicz}
\author[inst2]{Nicholson Koukpaizan}
\author[inst3]{Tobias Ribizel}
\author[inst3]{Fritz G\"{o}bel}
\author[inst1]{Shrirang Abhyankar}
\author[inst4,inst3]{Hartwig Anzt}
\author[inst2]{Slaven Pele\v{s}}

\affiliation[inst1]{organization={Pacific Northwest National Laboratory},
            city={Richland},
            postcode={99352}, 
            state={WA},
            country={USA}}

\affiliation[inst2]{organization={Oak Ridge National Laboratory},
            addressline={1 Bethel Valley Rdo}, 
            city={Oak Ridge},
            postcode={37830}, 
            state={TN},
            country={USA}}
\affiliation[inst3]{organization={Karlsruhe Institute of Technology},
            addressline={Kaiserstraße 12}, 
            city={Karlsruhe},
            postcode={76131}, 
            state={BW},
            country={Germany}}
\affiliation[inst4]{organization={University of Tennessee},
            addressline={203 Claxton Complex}, 
            city={Knoxville},
            postcode={37996}, 
            state={TN},
            country={USA}}

\begin{abstract}
Integrating renewable resources within the transmission grid at a wide scale poses significant challenges for economic dispatch as it requires analysis with more optimization parameters, constraints, and sources of uncertainty. 
This motivates the investigation of more efficient computational methods, especially those for solving the underlying linear systems, which typically take more than half of the overall computation time. In this paper, we present our work on sparse linear solvers that take advantage of hardware accelerators, such as \acp{gpu}, and improve the overall performance when used within economic dispatch computations. 
We treat the problems as sparse, which allows for faster execution but also makes the implementation of numerical methods more challenging.
We present the first GPU-native sparse direct solver that can execute on both AMD and NVIDIA GPUs.
We demonstrate significant performance improvements when using high-performance linear solvers within \ac{acopf} analysis.
Furthermore, we demonstrate the feasibility of getting significant performance improvements by executing the entire computation on \ac{gpu}-based hardware. Finally, we identify outstanding research issues and opportunities for even better utilization of heterogeneous systems, including those equipped with \acp{gpu}.
\end{abstract}








\end{frontmatter}


\section{Introduction}

Power grid utilities heavily rely on simulation and analysis tools for near- and long-term planning. With increased energy needs, more variable generation added to the grid, less predictable weather patterns, and ever-increasing cyber threats, the electric grid planning and operation have become more complex and compute-intensive. In particular, the \ac{acopf} \cite{ONeill2012, Frank2012} analysis, which is at the core of economic dispatch, has become computationally more demanding, pushing the limits of existing tools. New heterogeneous computing technologies utilizing \acp{gpu}, provide affordable high-density computational power that can address emerging needs in the energy industry. However, mathematical and computational methods typically used for economic dispatch are not designed for and perform poorly on \acp{gpu}. This is particularly true when solving underlying linear systems \cite{swirydowicz2022linear}, which typically makes up more than half of the overall computational cost.

 \acp{gpu} deliver high performance by executing groups of threads (warps) within single instruction sequence control. In \nvidia GPUs, $32$ threads form a warp. In AMD GPUs, $64$ threads form a wavefront -- which is equivalent to a warp. Standard \acp{cpu} have separate instruction sequence control for each thread. This means that the computational power of a \ac{gpu} can be harnessed only if the computation is expressed in terms of \ac{simd} operations. Hardware accelerators also have smaller level-1 cache memory compared to \acp{cpu}, which means that poor data coalescence in device memory can severely degrade computational performance. This also means that moving data in hardware accelerator memory is more expensive than moving data in conventional \ac{cpu} memory. These constraints need to be factored in when developing mathematical algorithms for execution on heterogeneous hardware.

There have been several efforts to develop \ac{gpu}-accelerated sparse linear solvers that are effective for computations in the power systems domain, including electromagnetic transient~\cite{dinkelbach2021factorisation, razik2019comparative} and power flow \cite{dorto2021comparing} simulations. There is far less reporting on linear solvers suitable for \ac{acopf} analysis due to the special properties of linear systems there. Existing results pertain mainly to dense formulations of the underlying linear systems \cite{Rakai2014,abhyankar2021acopf}.
 However, the computational complexity of dense linear solvers is $O(N^3)$ and their memory complexity is $O(N^2)$, where $N$ is the number of linear equations. For larger grid models, all performance gains on \ac{gpu} are offset by a cubic increase in computational cost and a quadratic increase in memory as shown in, e.g., \cite{su2020full}. 

In this paper, we discuss the solution of \ac{acopf} on \ac{gpu}s using sparse linear solvers. The main contributions of this paper are:
\begin{itemize}
\item Two refactorization-based sparse linear solvers -- one developed using CUDA libraries \cite{nvidia2021cusolver} and the other developed within open-source \gko library \cite{anzt2022ginkgo}. { Both} outperform existing \ac{gpu}-enabled solvers by a wide margin when used within \ac{acopf} analysis. 
\item { Demonstration} of meaningful speedup when our \ac{gpu} solvers are used in \ac{acopf} compared to the state-of-the-art baseline on \ac{cpu} hardware. We observed matrix factorization speedup of $2.4$--$3.4\times$ and triangular solve speedup of 3.2--$5.8\times$, which projects to $1.6$--$1.9\times$ overall speedup for the entire \ac{acopf} analysis compared to the \ac{cpu} baseline.
\item Performance profiles of \ac{acopf} analysis on NVIDIA and AMD \acp{gpu}, as well as on \ac{cpu} platforms. We identify key performance bottlenecks and estimate additional performance improvement attainable with currently existing technologies. 
\end{itemize}

The paper is organized as follows: In Section \ref{sec:acopf}, we describe properties of linear systems arising in \ac{acopf} analysis. In Section \ref{sec:testbed}, we describe the environment and use cases for testing the performance of linear solvers, and establish the state-of-the-art baseline. We discuss challenges when solving those linear systems on hardware accelerators in Section \ref{sec:gpusolvers}, and present our results in Section \ref{sec:results}. Finally, in Section \ref{sec:conclusion}, we summarize our findings and discuss the follow-on research that will lead to further performance improvements on heterogeneous hardware.

\section{Problem Definition}
\label{sec:acopf}

We test our linear solvers within an \ac{acopf} analysis, which is representative of optimization computations used in economic dispatch.
Mathematically, an \ac{acopf} is posed as a nonlinear non{-}convex optimization problem
\begin{subequations}\label{eq:problemstatement}
\begin{align}
          \min_x &~ F(x)               & \label{eq:opfobj} \\
  \text{s.t.}~~  & g(x) = 0,           & \label{eq:opfeq}  \\
           h^{-} & \le h(x) \le h^{+}, & \label{eq:opfineq}\\
           x^{-} & \le x    \le x^{+}, & \label{eq:opfbounds}
\end{align}
\end{subequations}
where $x$ is a vector of optimization variables, such as bus voltages and generator power outputs; vectors $x^{-},x^{+}$ define variable bounds, e.g.~resource capacity limits; $F(x)$ is a scalar objective function accounting for generator fuel, power imbalance, wind curtailment, and other power generation costs; $g(x)$ is a vector function that defines equality constraints, such as power balance; vector function $h(x)$ defines security constraints and vectors $h^{-},h^{+}$ specify security limits. For a more comprehensive discussion of the model aspects of the AC optimal power flow formulation, we refer the reader to {\cite{ONeill2012,Zimmerman2011}}.

We use interior method \cite{wachter2006implementation} 
to solve (\ref{eq:problemstatement}). 
To keep the presentation streamlined, we recast (\ref{eq:problemstatement}), without loss of generality, in a compact form
\begin{subequations}\label{problemstatement}
\begin{align}
   && \min_{y\in\mathbb{R}^{n}}\ \ & f(y)    \label{problemstatement_a} \\
   && \text{s.t.}        \ \ & c(y) = 0,    && \label{problemstatement_b} \\
   &&                        &    y \ge 0,  && \label{problemstatement_c}  
\end{align}
\end{subequations}
where inequality constraints (\ref{eq:opfineq}) are expressed as equality constraints $h(x) - h^{\pm} \pm s^{\pm} = 0$ using slack variables $s^-,s^+ \ge 0$. Similarly, bounds (\ref{eq:opfbounds}) are expressed in form (\ref{problemstatement_c}) by introducing variables $u^-,u^+ \ge 0$ such that $u^{\pm} = \mp(x - x^{\pm})$. The vector of optimization primal variables is $y=(u^-,u^+,s^-,s^+) \in \mathbb{R}^{n}$ and $f:~\mathbb{R}^{n}\rightarrow \mathbb{R}$ is a possibly non-convex objective function. All constraints are formulated in $c: \mathbb{R}^n \to \mathbb{R}^m$, where $m$ is the number of equality and inequality constraints.

Interior methods enforce bound constraints (\ref{problemstatement_c}) by adding barrier functions to the objective (\ref{problemstatement_a})
$$ 
\min_{y\in\mathbb{R}^{n}}
 f(y) - \mu\sum_{j=1}^{n} \ln{y_j},
$$
where $\mu>0$ is the barrier parameter. Interior methods are most effective when exact first and second derivatives are available, as we assume for $f(y)$ and $c(y)$. We define
$J(y) = \nabla c(y)$ as the sparse $m \times n$ Jacobian matrix for the constraints. 
The solution of a barrier subproblem satisfies the nonlinear equations
\begin{subequations}\label{eq:nonlinearequations}
\begin{align}
  \nabla f(y) + J(y) \lambda - z &= 0,  \label{nonlinearequations_a} \\ 
                            c(y) &= 0,  \label{nonlinearequations_c} \\
                             Y z &= \mu e, \label{nonlinearequations_e} 
\end{align}
\end{subequations}
where $\lambda\in \mathbb{R}^{m}$ is a vector of Lagrange multipliers (dual variables) for constraints (\ref{problemstatement_b}), $z$ are Lagrange multipliers for the bounds (\ref{problemstatement_c}), the matrix $Y\equiv \mathrm{diag}(y)$, and $e$ is $n$-dimensional vector with all elements equal to one. 

The barrier subproblem (\ref{eq:nonlinearequations}) is solved using a variant of a Newton method, and the solution to the original system (\ref{eq:problemstatement}) is recovered by a continuation method driving $\mu \to 0$. As it searches for the optimal solution, the interior method solves a sequence of the linearized barrier subproblems $K_k \Delta x_k = r_k$, $k=1,2,\dots$, where index $k$ denotes optimization solver iteration (including continuation step in $\mu$ and Newton iterations). Note that for $\mu=0$, $K_k$ is singular. A typical implementation of the interior method eliminates the linearized version of (\ref{nonlinearequations_e}) by substituting it in the linearized version of (\ref{nonlinearequations_a}) to obtain a symmetric linear system of \ac{kkt} type \cite{wachter2006implementation}
\begin{align} \label{eq:kktlinear}
\overbrace{\begin{bmatrix}
      H + D_y & J
     \\ J^T   & 0 
  \end{bmatrix}}^{K_k}
  \overbrace{\begin{bmatrix}
    \Delta y \\ 
    \Delta \lambda
  \end{bmatrix}}^{\Delta x_k}=
  \overbrace{\begin{bmatrix}
    r_{y} \\ r_{\lambda}
  \end{bmatrix}}^{r_k},
\end{align}
where $\Delta x_k$ is a search direction and $r_k$ is residual of (\ref{eq:nonlinearequations}) evaluated at current values of primal and dual variables. 
The Hessian
\begin{equation*}
H \equiv \nabla^2 f(y) + \sum_{i=1}^{m} \lambda_{i} \nabla^2 c_i(y),
\end{equation*}
is a sparse symmetric $n \times n$ matrix and 
$D_y \equiv \mu Y^{-2}$ is a diagonal  $n \times n$ matrix.

The \ac{kkt} linear system (\ref{eq:kktlinear}) is sparse symmetric indefinite and typically \textit{ill-conditioned}. In \ac{acopf} and other analyses for economic dispatch, $K_k$ matrices are \textit{extremely sparse} due to the connectivity structure of power grids. 
The interior method needs to return a solution after $\mu$ is small enough so the solution of (\ref{eq:nonlinearequations}) approximates the solution of (\ref{problemstatement}) well, but before $K_k$ in (\ref{eq:kktlinear}) becomes too ill-conditioned. Having a solver that can compute an accurate solution to ill-conditioned linear systems without significant performance penalty is critical for accurate and efficient \ac{acopf} analysis, especially since solving (\ref{eq:kktlinear}) is the major part of the overall computational cost.

We note that all $K_k$ in (\ref{eq:kktlinear}) have the same sparsity pattern. An efficient linear solver shall take advantage of constant sparsity patterns and reuse parts of computations over the sequence of $K_k$ matrices when possible.

\section{Test Networks and Optimization Setup}
\label{sec:testbed}

As our test cases, we use synthetic grid models: Northeastern, Eastern, and combined Western and Eastern U.S. grid \cite{birchfield2017tamu-cases}. The details of our test cases are provided in Table \ref{tab:descr}.
\begin{table}[h!] 
\caption{Characteristics of the three test networks specifying the number of buses, generators, and lines. The specifics of the linear system (\ref{eq:kktlinear}) for each of these networks are given in terms of the matrix size (N) and number of non-zeros (nnz) for the matrix $K_k$. Numbers are rounded to 3 digits. K and M denote $10^3$ and $10^6$, respectively.}
\label{tab:descr}

\centering

\scalebox{0.68}{
\begin{tabular}{lrrrrr} \toprule
   Grid    & Buses & Generators & Lines    & $N(K_k)$ & nnz($K_k$) 
\\ \midrule
   Northeastern US        & 25 K &  4.8 K &  32.3 K & 108 K  & 1.19 M
\\ Eastern US             & 70 K & 10.4 K &  88.2 K & 296 K  & 3.20 M 
\\ Western and Eastern US & 80 K & 13.4 K & 104.1 K & 340 K  & 3.73 M
\\ \bottomrule
\end{tabular}
}
\end{table}

\begin{figure}[ht]
\centering
  {\includegraphics[width=\columnwidth,trim={0cm 0cm 0cm 0cm},clip]{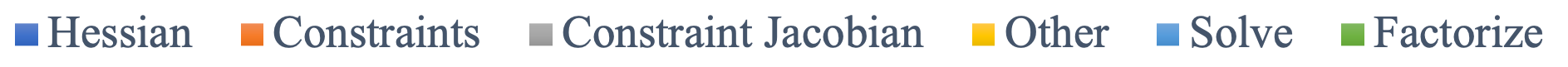}} \\
  \begin{subfigure}{0.4\columnwidth}
    {\includegraphics[width=\columnwidth,trim={3cm 2cm 3cm 0.1cm},clip]{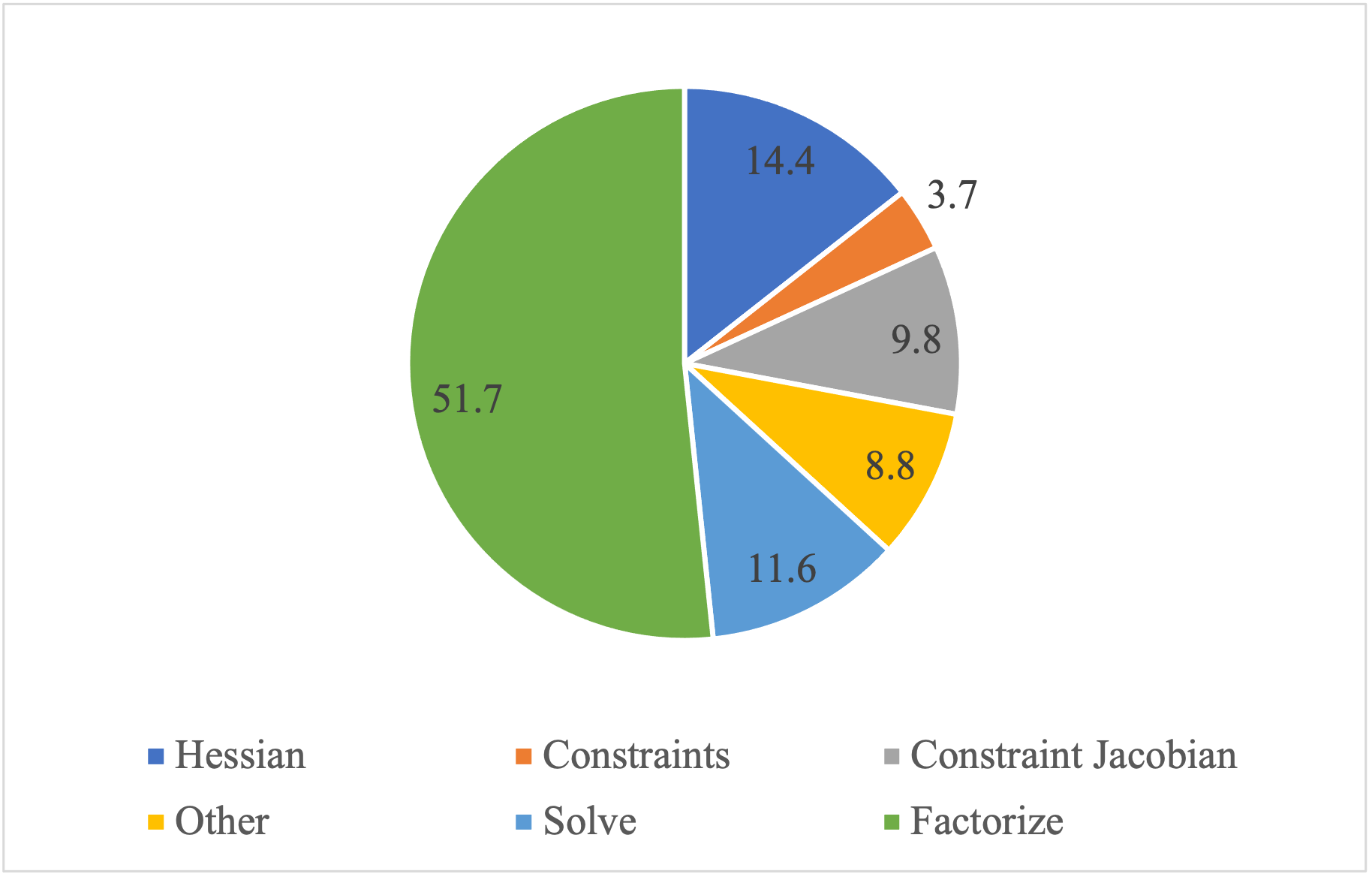}}
    \caption{Northeast U.S. grid}
  \end{subfigure}
  \begin{subfigure}{0.4\columnwidth}
    {\includegraphics[width=\columnwidth,trim={3cm 2cm 3cm 0.1cm},clip]{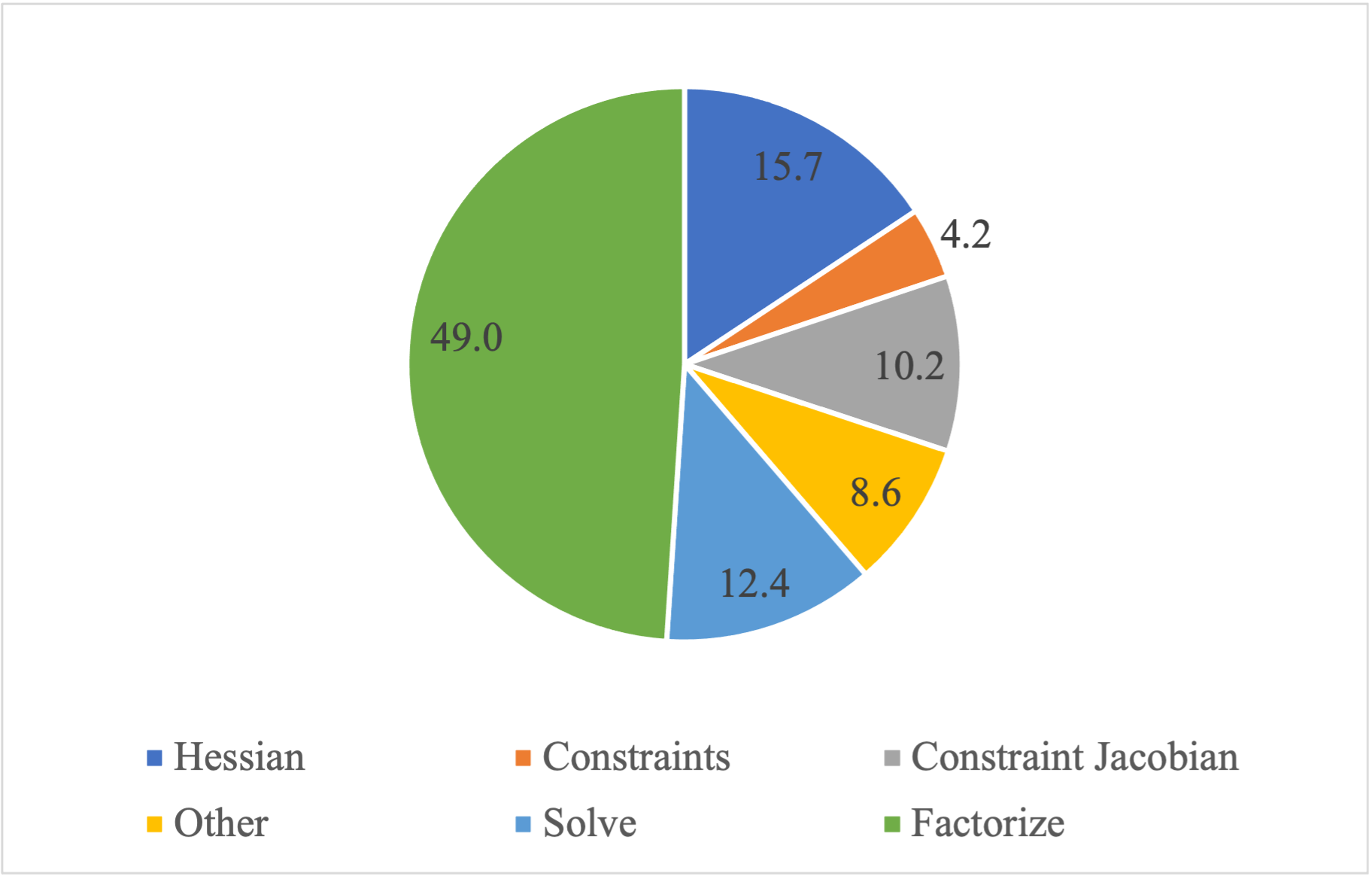}}
    \caption{Eastern U.S. grid}
  \end{subfigure}
  \begin{subfigure}{0.4\columnwidth}
    {\includegraphics[width=\columnwidth,trim={3cm 2cm 3cm 0.1cm},clip]{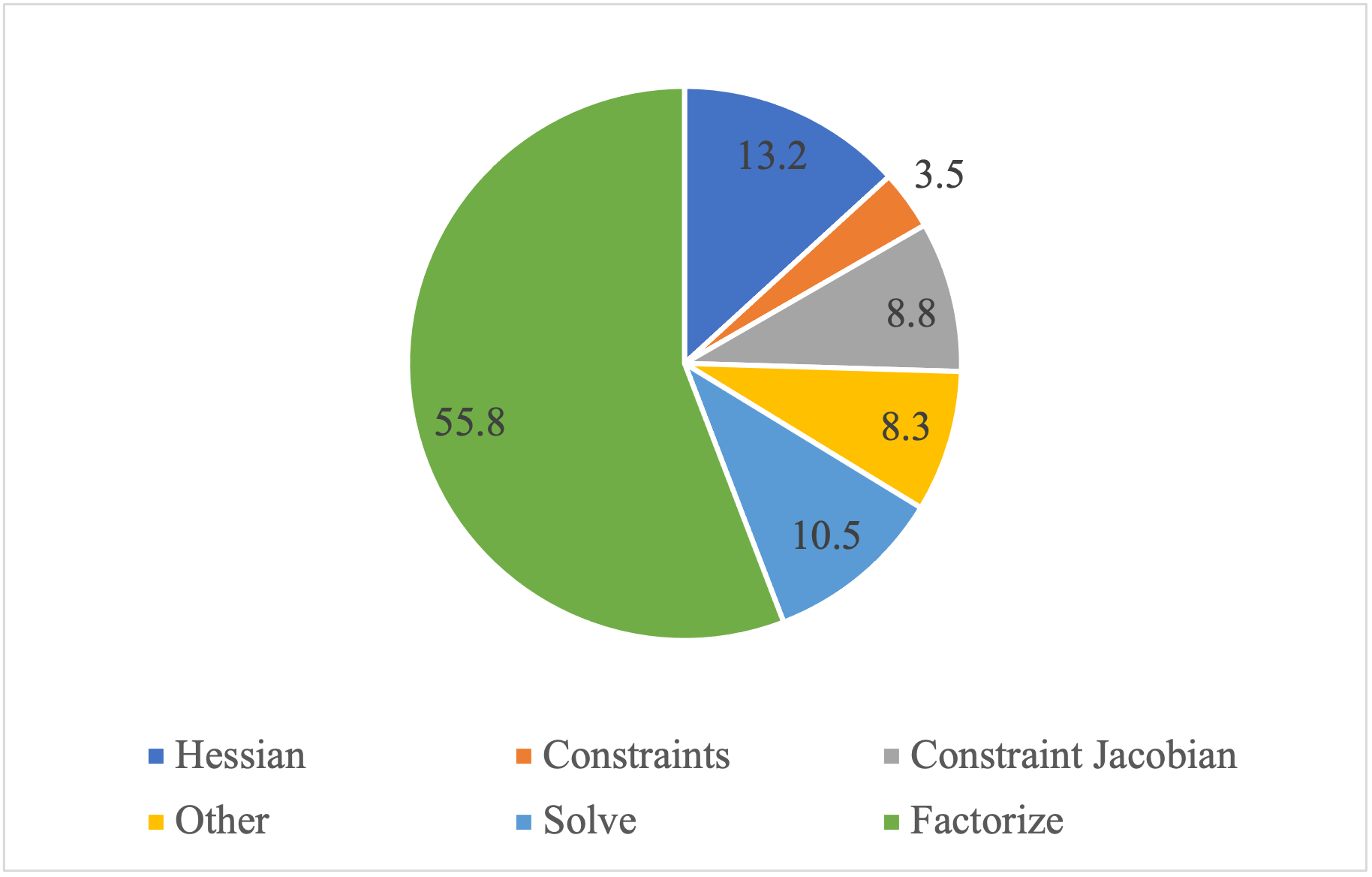}}
    \caption{Eastern and Western U.S. grids}
  \end{subfigure}

  \caption{Computational cost of ACOPF broken down by analysis functions. Linear solver functions (matrix factorization and triangular solve) contribute to 60\% of the overall ACOPF compute time.}
  \label{fig:ma57cost}
\end{figure}

To establish a baseline for evaluating our new solvers, we use MA57 linear solver \cite{duff2004ma57}, which is commonly used in commercial and open-source tools when solving \ac{acopf} and similar optimization problems. { When using MA57, symbolic factorization is performed only once for all systems with the same sparsity pattern.} 
The profiling results for the solution of the KKT system for these three networks with MA57 linear solver show that solving the linear system takes up more than 60\% of the overall \ac{acopf} cost (see Figure \ref{fig:ma57cost}). { These results were obtained on an IBM Power 9 \ac{cpu}; the computing environment is described in more detail in section \ref{subsec:profiling}.} Starting from this baseline, any meaningful speedup of \ac{acopf} requires accelerating the linear solver.

For the \ac{acopf} analysis we use \exago \cite{ExaGo}, an open-source package for solving large-scale optimization problems on CPU and GPU-based platforms involving security, stochastic, and multi-period constraints. \exago models the ACOPF problem in a power-balance network formulation with a minimum generation cost objective function and enforcing voltage, capacity, and line flow constraints. It uses the open-source
\hiop optimization engine \cite{hiop_techrep} to solve the resultant optimization problem. To our knowledge, \hiop is the only open-source optimization solver capable of solving nonlinear optimization problems on GPUs. The developed linear solvers described in this paper are interfaced to and called from \hiop. The \ac{acopf} application provided with \exago is used to configure and run our tests.
\exago parses input data files for power grids, generates model objects $H$, $J$, $D_y$, $r_y$, and $r_{\lambda}$, sets variable limits, and passes them to \hiop. The optimization engine assembles (\ref{eq:kktlinear}), updates it at each iteration, and passes it to the linear solver. The solution $\Delta x_k$ to (\ref{eq:kktlinear}) is used by \hiop to update the vector of primal variables in \exago for the next iteration. {In our computations we set relative tolerance for the interior method in \hiop to $10^{-6}$. This leads to barrier parameter values reduced to $\mu \sim 10^{-7}$ before converged solution is obtained.}

A typical implementation of the interior method requires a linear solver to provide matrix inertia (number of positive, negative, and zero matrix eigenvalues). Matrix inertia can be readily obtained in $LDL^T$, but not in $LU$ factorization. It is important to note that \hiop offers an inertia-free interior method implementation \cite{chiang2016inertia}, as an option, which allows us to use $LU$ linear solvers for \ac{acopf} analysis.

\section{GPU-enabled Linear Solvers}
\label{sec:gpusolvers}

There exists a vast body of literature on direct linear solvers for heterogeneous compute architectures~\cite{li2003superlu_dist, ghysels2016efficient}. Moreover, the recent decade brought increased interest in developing solvers that are capable of using the \acp{gpu} efficiently~\cite{hogg2016sparse}. These efforts, however, usually target {\it generic} problems that are well-conditioned and characterized by a well-defined structure (i.e., matrices that contain dense blocks of entries). The available parallel techniques rely on multifrontal or supernodal approaches; in both cases, dense blocks within the matrix are essential for satisfactory performance on the \acp{gpu}. In \cite{swirydowicz2022linear} we identified and tested five applicable linear solver packages: SuperLU~\cite{li05}, STRUMPACK~\cite{ghysels2016efficient}, SPRAL-SSIDS~\cite{Duff2020}, PaStiX~\cite{henon2002pastix} and \verb|cuSolverSP| (we tested the {\it black-box} QR and LU-based solve functions). The results were compared to the MA57~\cite{duff2004ma57}, which is a single thread and \ac{cpu}-only code. The testing revealed that none of the GPU-accelerated packages was substantially better than MA57. Moreover, turning on GPU acceleration within each package often resulted in performance losses when compared to the \ac{cpu}-only code. The timing data published in \cite{swirydowicz2022linear} is summarized in Table~\ref{tab:old_gpu_cpu}. The presented timing results are averaged over a few representative matrix systems for each solver. 

\begin{table*}[htb]
\centering
\scalebox{0.8}{
\begin{tabular}{ccccccccc}
\toprule
\multicolumn{1}{c}{\textbf{Test case}} & \multicolumn{1}{c}{\textbf{Size}} & \multicolumn{1}{c}{\textbf{NNZ}} & \multicolumn{1}{c}{} & \multicolumn{1}{c}{\textbf{SuperLU}} & \multicolumn{1}{c}{\textbf{STRUMPACK}} & \multicolumn{1}{c}{\textbf{cuSolver QR}} & \multicolumn{1}{c}{\textbf{SSIDS}} & \multicolumn{1}{c}{\textbf{MA57}}\\
\midrule
\multirow{2}{*}{\textbf{Case 1}} & \multirow{2}{*}{55.7K}& \multirow{2}{*}{268K}& CPU (s) & 1.1 & 1.0 & 2.6 & 0.7 & 0.2 \\
&&& GPU (s) & 1.6 & 1.6 & 1.8 & 5.1 & -- \\
\midrule
\multirow{2}{*}{\textbf{Case 2}} & \multirow{2}{*}{238K}& \multirow{2}{*}{1.11M}&CPU (s) & 4.0 & 2.8 & 18.2 & 2.6 & 0.8 \\
&&& GPU (s) & 5.0 & 3.7 & 5.7 & 4.8 & -- \\
\midrule
\multirow{2}{*}{\textbf{Case 3}} &\multirow{2}{*}{296K}&\multirow{2}{*}{7.67M}&CPU (s) & 30 & 24 & 614 & 29 & 6 \\
&&& GPU (s) & 33 & 24 & -- & 198 & -- \\
\bottomrule
\end{tabular}}
\caption{Test results with various GPU-accelerated linear solvers for representative systems generated with Ipopt~\cite{swirydowicz2022linear}. The test linear systems are available at \href{https://github.com/NREL/opf_matrices}{https://github.com/NREL/opf\_matrices}. We used the last $3$ systems for test case 1 (ACTIVSg2000 in the repository) and the last $5$ for the two remaining test cases (ACTIVSg10k and ACTIVSg70k in the repository). \label{tab:old_gpu_cpu} }
\end{table*}



The development of GPU-resident solvers for power flow problems is challenging for several reasons. First of all, it is the nature of sparse direct solvers that allows only for a limited amount of parallelism. The situation becomes worse if the sparse matrix has no inherent block structure, which is the case of power flow problems. {As mentioned earlier in this section,} in this case, supervariable agglomeration and multifrontal approaches that accumulate elements in dense blocks and invoke dense linear algebra routines are not suitable (which explains the results from \cite{swirydowicz2022linear}). Instead, we require fine-grained scheduling of individual variable eliminations once their dependencies have completed processing. Finally, the power grid problems require pivoting for numerical stability.  Pivoting is a bottleneck on GPUs as it requires synchronization and inter-block communication, further degrading the performance.

From a high-level standpoint, {sparse} direct solver computations can be split into three phases: ($i$) symbolic factorization, when the matrix is reordered and the structure of $L$ and $U$ factors are set, together with permutation order; ($ii$) numeric phase, when numeric factorization is computed; and ($iii$) solve phase, during which the factors obtained in the previous phase are used to compute the solution. 

The symbolic phase of the computation is typically done on the \ac{cpu}, sometimes with marginal \ac{gpu} offloading. The symbolic factorization does rely on the non-zero structure of the matrix and not on its values. In the case of the KKT matrices, as mentioned before, the non-zero structure of the matrix systems does not change during the optimization solver run, and hence the symbolic phase (or parts of it) can be executed once and reused for all relevant systems. This idea ties in with the idea of {\it refactorization}.
This term is used somewhat loosely in the literature; it usually means that once {the permutation vectors and non-zero patterns of triangular factors were computed in the \textit{numerical} factorization, they are reused for the next system in the sequence. } Refactorization is implemented in some of the existing direct solvers, for instance, in KLU~\cite{davis2010algorithm}, NICSLU~\cite{chen2013nicslu}, \nvidia's \verb|cuSolverRf|, and undocumented but publicly released \nvidia's \verb|cuSolverGLU|. Refactorization has not been particularly popular in sparse direct \ac{cpu} solvers as the cost of pivoting on the \ac{cpu} is generally not a concern. In contrast, pivoting on the \ac{gpu} is prohibitively expensive, and avoiding pivoting is paramount for  \ac{gpu} speedups. 
%

We found that refactorization approach is very effective for linear systems arising in \ac{acopf} (see Section~\ref{sec:results}). Similar approaches were successfully used {before} in modeling power systems.
For instance, \cite{dinkelbach2021factorisation} develops a partial refactorization method (on the \ac{cpu}, within NICSLU package) and compares its efficiency to KLU, SuperLU, and non-altered NICSLU. The authors use standalone matrices obtained from dynamic phasor simulations as their test cases. In \cite{razik2019comparative}, the authors compare the performance of NICSLU and KLU (both with refactorization) and GLU~\cite{he2015gpu}, which is a sparse direct solver that performs factorization on the \ac{gpu} (but does not enable refactorization); the approach was also tested on matrices extracted from dynamic phasor simulations and the results favored NICSLU. Six different approaches for power flow simulations are tested in \cite{dorto2021comparing}, two of them being \ac{gpu}-only, two \ac{cpu}-only and two hybrid (\ac{gpu}+\ac{cpu}). The tested \ac{gpu} approaches involve refactorization based on \verb|cuSolverGLU|. The hybrid solver turns out to be the fastest (i.e., preprocessing and symbolic factorization are performed on the \ac{cpu}, the refactorization happens on the \ac{gpu}, the triangular solve is done on the \ac{cpu}).

\section{Approach and Results}
\label{sec:results}

We present two refactorization approaches and demonstrate their performance in the \ac{acopf} analysis context. The first approach we developed uses KLU and proprietary \verb|cuSolver| libraries. The second approach is developed within the open-source \gko library. Both are hybrid \ac{cpu}-\ac{gpu} approaches that exploit the property that the matrix non-zero structure in (\ref{eq:kktlinear}) does not change from one system to the next {so the same pivot sequence can be reused. Both approaches also take advantage of \ac{kkt} matrix regularization performed by the interior method \cite{wachter2006implementation}, which has a similar effect as perturbation techniques used in static pivoting \cite{duff2007static}, and helps us reuse the same pivot sequence over a larger number of linear systems}. In contrast to the hybrid approach in \cite{dorto2021comparing}, we move linear system data to \ac{gpu} only once, after solving the first system in (\ref{eq:kktlinear}), and perform all subsequent computations there. In this way, we avoid excessive data movement between \ac{cpu} and \ac{gpu}, which adds significant overhead to the overall computational time.

{
\subsection{Refactorization Solver Using CUDA Libraries}

\subsubsection{Approach}
}
In this approach, we completely solve the first system on the \ac{cpu} using KLU with approximate minimum degree (AMD) reordering \cite{AMD_reordering}. Then, we extract the elements of the symbolic factorization (i.e., the permutation vectors and the sparse matrix structures of the factors). Next, we set up an appropriate \verb|cuSolver| (Rf or GLU) data structure, and we then solve all the remaining systems at each next optimization solver step by calling refactorization function. Typically, hundreds of linear systems need to be solved during \ac{acopf} execution, hence the cost of solving the first system(s) on the \ac{cpu} is amortized over many optimization solver steps. Both \verb|cuSolverRf| and \verb|cuSolverGLU| allow the user to provide permutation vectors and the sparse non-zero structure of $L$ and $U$ matrices obtained by the LU factorization of choice. We tested the native \verb|cuSolver| factorization methods on \ac{cpu}, however, we obtained the best results with KLU (this agrees with the results from~\cite{dorto2021comparing}). KLU also produced the sparsest triangular factors. 
Since {the same pivot sequence is used over and over} in refactorization, the computed solution {could} potentially have a large error. Many (but not all) linear solvers follow the factorization by iterative refinement~\cite{wilkinson1965rounding}, which is a method of improving the backward error in the solution. For instance, MA57 {by default} uses up to $10$ iterations of iterative refinement; for the test matrices from ~\cite{swirydowicz2022linear} the average was around $6$. 
How does this work? Let us assume that we solved the linear system (\ref{eq:kktlinear}) using a direct method and computed a solution, $\Delta x_k^{(0)}$. To improve the solution, $\rho^{(0)} = r_k-K_k \Delta x_k^{(0)}$ is computed, and a new system is solved for $\delta^{(0)}$: $K_k \delta^{(0)} =\rho^{(0)}$ using the factors of $K_k$ obtained previously. Then, a new solution is formed $\Delta x_k^{(1)} = \Delta x_k^{(0)}+\delta^{(0)}$. If the solution is still not satisfactory, $\rho^{(1)} = r_k - K_k \Delta x_k^{(1)}$ is computed and the process repeats. This method is often quite effective. An alternative method is to use the triangular factors as a preconditioner inside the linear solver in the same way one would use the incomplete LU (ILU) preconditioner \cite{saad2003iterative}. The literature suggests using flexible GMRES \cite{saad1993flexible} as the iterative method deeming it more stable in this case~\cite{arioli2007note}. Some authors, e.g.~\cite{carson2020three}, mix the two approaches by using an iterative solver to solve the systems $K_k \delta^{(i)} =\rho^{(i)}$. We use the {\it ILU style approach} because in practice it requires fewer triangular solves and leads to a solution with the same quality. 

{ Another method to address the deterioration in the solution quality is to recompute the symbolic factorization (on the CPU) if the solution quality becomes too poor or if the maximum number of refinement iterations is exceeded. In our test cases, however, we typically needed only one or two refinement iterations to maintain an appropriate level of accuracy. Hence, we never recomputed the symbolic factorization. }

{
\subsubsection{Implementation}
}

The refactorization approach is implemented in CUDA/C++ and currently is distributed with \hiop library. The algorithm follows the steps below.

\textbf{Solve the first system in the sequence.} The first system in the sequence (\ref{eq:kktlinear}) is solved completely using KLU (using the usual three-step approach: symbolic factorization, numeric factorization and solve). We use AMD reordering and we allow the factorization to continue even if the matrix is deemed to be singular by the analysis (which is one of the user-defined KLU options).

\textbf{Set the sparsity pattern of the factors.} The non-zero structures of the $L$ and $U$ factors are extracted from KLU together with permutation vectors. The factors are returned in compressed sparse column (CSC) format. If \verb|cuSolverGLU| is used, the factors are combined into a matrix containing both $L$ and $U$ and the matrix is copied to the \ac{gpu}. If \verb|cuSolverRf| is used, both factors are copied to the \ac{gpu} separately and each factor is converted to compressed sparse row (CSR) format on the \ac{gpu}. 

\textbf{Refactorization setup.} An appropriate \verb|cuSolver| refactorization setup and analysis functions are called (there are different functions for \verb|cuSolverRf| and for \verb|cuSolverGLU|). 

\textbf{Solve subsequent systems.} Once the refactorization data is set up, each subsequent system is solved on the \ac{gpu}. This consists of two functions: resetting the values inside the refactorization data structure and performing the refactorization (factorizing the new matrix {using the prior pivot sequence}). Each refactorization method has its own functions to execute these operations. \verb|cuSolverRf| follows up with a ``solve'' function that permutes the solution to the original order and performs two triangular solves. \verb|cuSolverGLU| also adds MA57-style iterative refinement inside its black-box ``solve'' function. { In the case of \verb|cuSolverRf|, we  investigated an approach as in~\cite{dorto2021comparing}, in which the triangular solve happens on the \ac{cpu}. However, for our test cases, the GPU-resident \verb|cuSolverRf| triangular solver was overall faster, as it avoids moving matrix factors and permutation vectors to \ac{cpu} and solution vector back to \ac{gpu} each time the triangular solver is called. }

\textbf{Iterative refinement.} If desired (or if the error in the solution is too large), \verb|cuSolverRf| solve can be followed by iterative refinement. We implemented it as flexible GMRES with re-orthogonalized classical Gram-Schmidt (CGS2) orthogonalization (to improve numerical stability and \ac{gpu} performance). 

The iterative refinement was implemented to be highly efficient on the \ac{gpu}. For instance, all the large arrays (such as the ones used to store Krylov vectors) are allocated once and reused for all the systems. The handles (needed for various CUDA libraries, such as \verb|cuBlas|) and buffers (such as the one used in the matrix-vector product) are also allocated once and reused. For most of the test cases only one or two steps of iterative refinement were needed.

Both \verb|cuSolver| refactorization solvers come with limitations. The faster out of the two, \verb|cuSolverGLU| does not provide access to the triangular factors, which makes it impossible to couple it with alternative (faster) sparse triangular solvers or user-implemented iterative refinement. \verb|cuSolverGLU| also comes with its own iterative refinement (part of the {\it solve} function) which cannot be turned off and is not parameterized (i.e., the user cannot specify { the} maximum number of iterations, the algorithm used, or the solver tolerance). Both refactorization solvers are closed source, which makes it difficult to extract pivoting information and impossible to apply problem-specific optimization. Also, both  of them can only execute on \nvidia GPUs, which limits the platform support.

Against this background, we explored the possibility to design and deploy platform-portable open-source direct solver functionality in the \gko math library.

\subsection{\gko Refactorization Solvers}

{ 
\subsubsection{Approach}

For the refactorization approach, we use the open-source software library
\gko that is developed within the Exascale Computing
Project (ECP)~\cite{anzt2022ginkgo}. 
\gko focuses on the efficient handling of sparse linear
systems on GPUs. Implemented in C++ and featuring backends in the
hardware-native languages CUDA (for NVIDIA GPUs), HIP (for AMD GPUs),
and DPC++ (for Intel GPUs), the library combines sustainability with
performance portability, see Figure \ref{ha:fig:ginkgo}. 

\begin{figure*}[h!]
\centering
\includegraphics[width=0.75\textwidth]{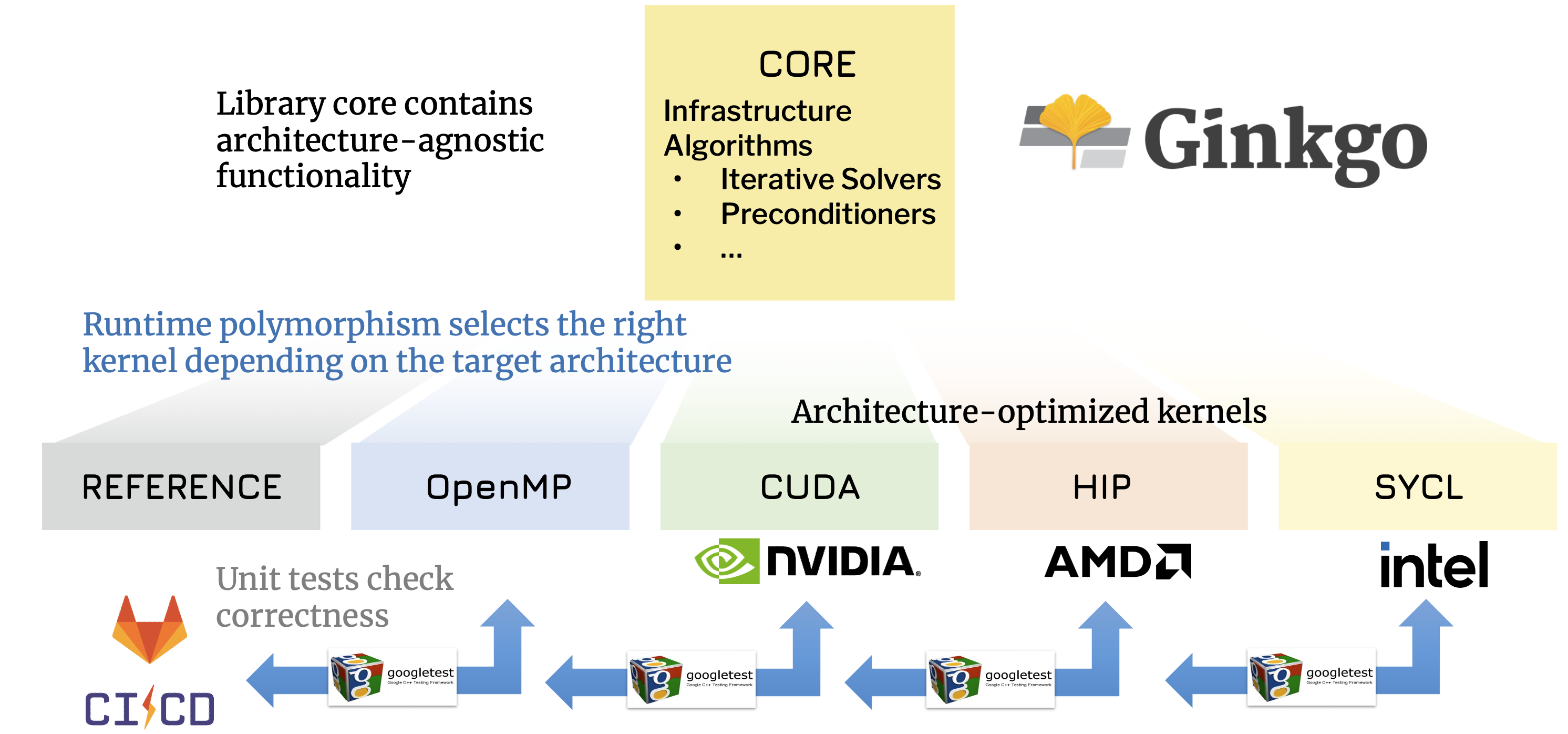}
\caption{The overall structure of the Ginkgo math library that was designed for performance portability.}
\label{ha:fig:ginkgo}
\end{figure*}

The implementation using the \gko framework consists of three phases: preprocessing, (re)factorization, and triangular solve. The preprocessing step is executed only once, and on the CPU, while the repeated calculations in the factorization happen entirely on the GPU. We preprocess the system matrix to improve numerical stability (pivoting and equilibration using the \texttt{MC64} algorithm introduced by Duff and Koster~\cite{mc64}) followed by a reordering to reduce fill-in (symmetric AMD on $A + A^T$). The resulting permutation and scaling factors will later be used to map the solution of the transformed system back to the original linear system.
Following the preprocessing, we compute a symbolic factorization of the scaled and reordered matrix. The resulting sparsity pattern of the $L$ and $U$ factors can be reused for all subsequent iterations of the optimization algorithm.
Finally, we compute the values of $L$ and $U$ on the GPU by first filling the values of $A$ into the combined storage for $L$ and $U$ and executing the numerical factorization kernel.}

\subsubsection{Implementation}
{ The components of the algorithm were implemented in the \gko library using C++ for the high-level control flow and CUDA/HIP for the factorization kernels. It is included in the Release 1.6.0 of \gko available on GitHub~\cite{ginkgo-release}.

\textbf{Preprocessing.} The reordering and equilibriation as well as the symbolic factorization of the resulting reordered matrix happens sequentially on the CPU. The symbolic $LU$ factorization closely follows the \texttt{fill1} algorithm by Rose and Tarjan~\cite{rose_algorithmic_1978}.}

\textbf{Numerical Factorization.} The $LU$ factorization kernel {(Algorithm~\ref{alg:ginkgo-lu})} uses a fine-grained up-looking parallel factorization based on the dependency structure encoded in the lower $L$ factor. The kernel relies on two important components: a sophisticated scheduling approach and an efficient sparse vector addition routine. Each row is mapped to a warp/wavefront on the \ac{gpu} and gets updated with all dependency rows. The updates add a sparse row from $U$ in-place to the current sparse row, thus eliminating the corresponding variable in the current row. This sparse vector addition is facilitated by a sparsity pattern lookup structure that uses bitmaps or hashtables, depending on the pattern, to compute the nonzero index to each column index in the row. The dependency resolution uses a sync-free scheduling approach based on a \emph{ready} flag for each row, {previously used for sparse triangular solvers by Lui et. al~\cite{liu2016syncfree}. The scheduling takes advantage of} \acp{gpu} from NVIDIA and AMD scheduling their thread blocks in monotonic order and providing strong forward-progress guarantees inside a single thread block.

\begin{algorithm}[htb]
\caption{{\gko LU factorization algorithm}}\label{alg:ginkgo-lu}
    \SetKwFor{ParFor}{parfor}{do}{end}
    set $\text{ready}[i] \leftarrow 0$ for all $i$\;
    \ParFor{\emph{row $i = 0, \dots, n - 1$}}{
        \For{\emph{lower nonzero} $l_{id}$}{
            \lWhile{\emph{ready[$d$] = 0}}{wait}
            $\alpha \leftarrow a_{id} / a_{ii}$\;
            $l_{ii} \leftarrow \alpha$\;
            \ParFor(\tcp*[h]{warp-parallel}){\emph{upper nonzero} $u_{dj}$}{
                $a_{ij} \leftarrow a_{ij} - \alpha \cdot u_{dj}$\;
            }
        }
        split $a_{i\cdot}$ into $l_{i\cdot}$ and $u_{i\cdot}$ \tcp*[l]{conceptually}
        $\text{ready}[i] \leftarrow 1$\;
    }
\end{algorithm}

{\textbf{Refactorization.}
All of the scaling coefficients and permutation indices from Step 1 and all of the lookup data structures and the symbolic factorization from Step 2 can be reused for subsequent linear systems, leaving only the numerical factorization itself to be recomputed.
Finally, we permute and scale the right-hand side and solution vectors into/out of a persistent workspace that needs only to be allocated once.

{ \textbf{Triangular Solvers.}} We solve the resulting triangular systems using the triangular solvers provided by the \verb|cuSPARSE| / \verb|rocSPARSE| library. Alternatively, the application can be set to use \gko's own triangular solvers based on a configuration option. These triangular solvers use the same sync-free scheduling approach as the factorization (Algorithm~\ref{alg:ginkgo-trisolve}). We use a thread-per-row mapping, which requires independent thread scheduling available on NVIDIA GPUs since the Volta architecture. On AMD GPUs and older NVIDIA GPUs, we need to modify the control flow to guarantee forward-progress for the entire warp.}

\begin{algorithm}[htb]
\caption{{\gko triangular solver algorithm solving $Lx = y$}}\label{alg:ginkgo-trisolve}
    \SetKwFor{ParFor}{parfor}{do}{end}
    set $x_i \leftarrow \text{NaN}$ for all $i$\;
    \ParFor{\emph{row $i = 0, \dots, n - 1$}}{
        $\tilde x \leftarrow y_i$\;
        \For{\emph{nonzero} $l_{id}$}{
            \lWhile{$x_d$ \emph{is NaN}}{wait}
            $\tilde x \leftarrow \tilde x - l_{id} x_d$\;
        }
        $x_i \leftarrow \tilde x$\;
    }
\end{algorithm}

\subsection{Performance Profiling}
\label{subsec:profiling}

We evaluated the approach described herein on the Summit supercomputer and the Crusher test and development system at the \ac{olcf}. Each Summit node is equipped with two 21-core IBM Power 9 \acp{cpu} and six \nvidia V100 \acp{gpu}, while Crusher's node architecture consists of one 64-core AMD EPYC 7A53 \ac{cpu} and four AMD MI250X \acp{gpu}. The AMD \ac{gpu} contains two \acp{gcd} that can be treated as individual \acp{gpu}. For the present benchmarks, we consider one \ac{cpu} core and one \ac{gpu}/\ac{gcd} (\nvidia V100 \ac{gpu} or AMD MI250X \ac{gcd}). The code was compiled with CUDA 11.4.2 and the GNU Compiler Collection 10.2.0 on the NVIDIA platform and with the ROCm 5.2.0 software stack on the AMD platform.

{We used state-of-the-art MA57 linear solver, which implements $LDL^T$ factorization, as our \ac{cpu} baseline. The MA57 was configured to perform symbolic factorization only once, for the first linear system in (\ref{eq:kktlinear}), and reuse it for all subsequent systems. Numerical factorization was configured to use default pivoting options and the triangular solve was configured to use no more than one iterative refinement iteration to maximize the performance. These are default MA57 settings in HiOp optimization library. We note it has been reported that performance of MA57 could be further improved if the numerical factorization is configured to use \textit{static pivoting} \cite{duff2007static} and triangular solve is followed with 10 or more refinement iterations to ensure sufficient solution accuracy is retained. When tested with our use cases, we found that static pivoting improved linear solver performance by 30\%. However, the solution quality deteriorated so much that even with up to 100 iterative refinement iterations, we could not recover solution accuracy needed for the optimization solver to converge.}

The MA57 and \gko solvers were tested on both systems, as \gko supports different \ac{gpu} backends \cite{gkoportability}, while the \verb|cuSolver| approach was only evaluated on NVIDIA \acp{gpu}. \verb|cuSolverRf| was evaluated with a maximum of $20$ refinement iterations and with the iterative refinement tolerance set to $10^{-14}$ (near the double floating point precision).

Tables \ref{tab:ma57_vs_cusolver_vs_ginkgo} and \ref{tab:ma57_vs_ginkgo} report the overall \ac{acopf} analysis performance when using different linear solvers for the test cases detailed in Table \ref{tab:descr}. { The reported total runtimes were each measured from a single profiling run. Multiple \ac{acopf} runs were subsequently performed to assess the performance variability, resulting in an observed population standard deviations (normalized by the mean value of the total runtime) lower than 1\%.} 

\begin{figure}[htb!]
\centering
  {\includegraphics[width=\columnwidth,trim={0cm 0cm 0cm 0cm},clip]{./profile_legend}} \\
  \begin{subfigure}{0.8\columnwidth}
    {\includegraphics[width=\columnwidth,trim={0.1cm 0.1cm 0.1cm 0.1cm},clip]{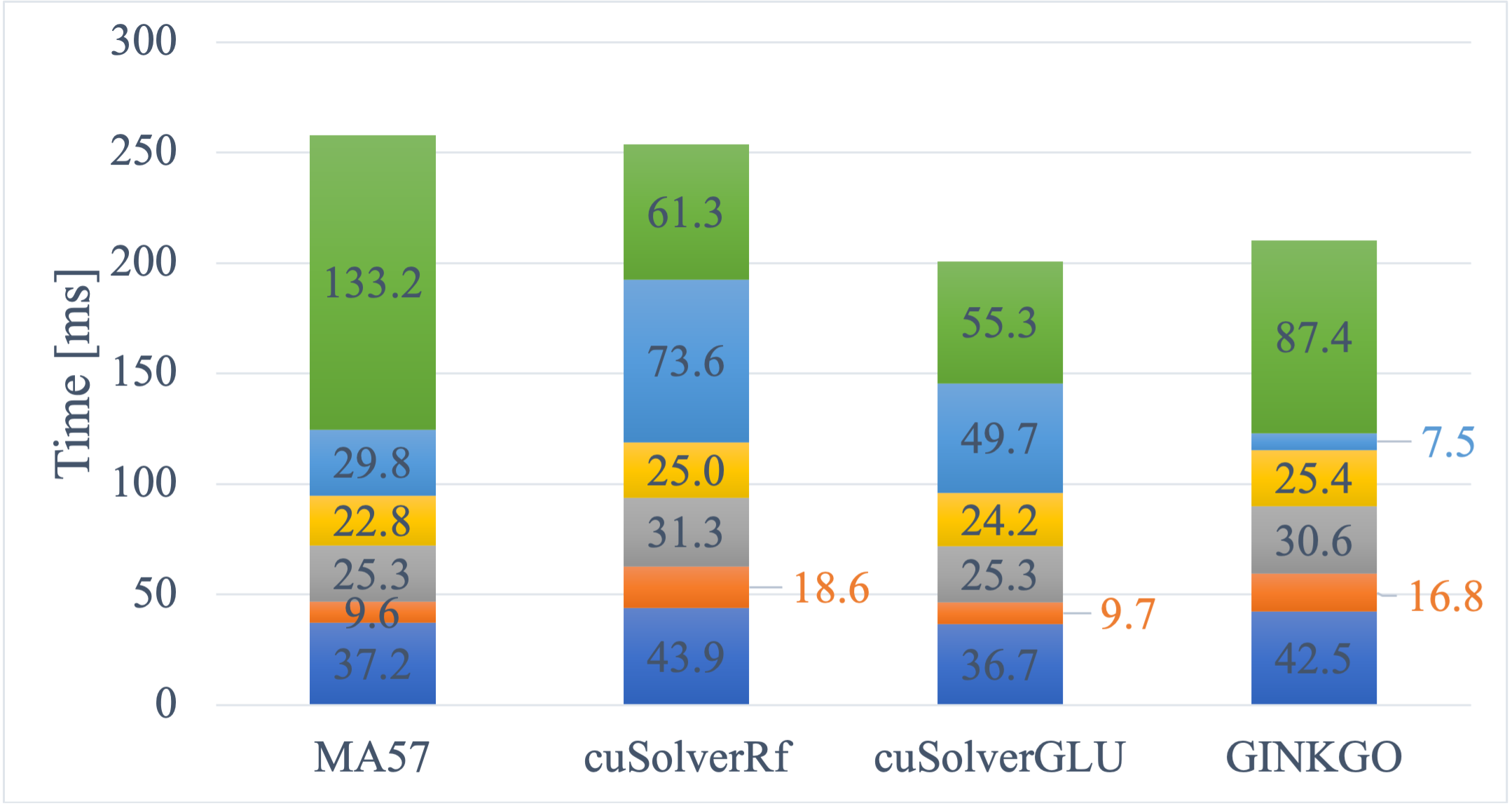}}
    \caption{Northeast U.S. grid}
  \end{subfigure}
  \begin{subfigure}{0.8\columnwidth}
    {\includegraphics[width=\columnwidth,trim={0.1cm 0.1cm 0.1cm 0.1cm},clip]{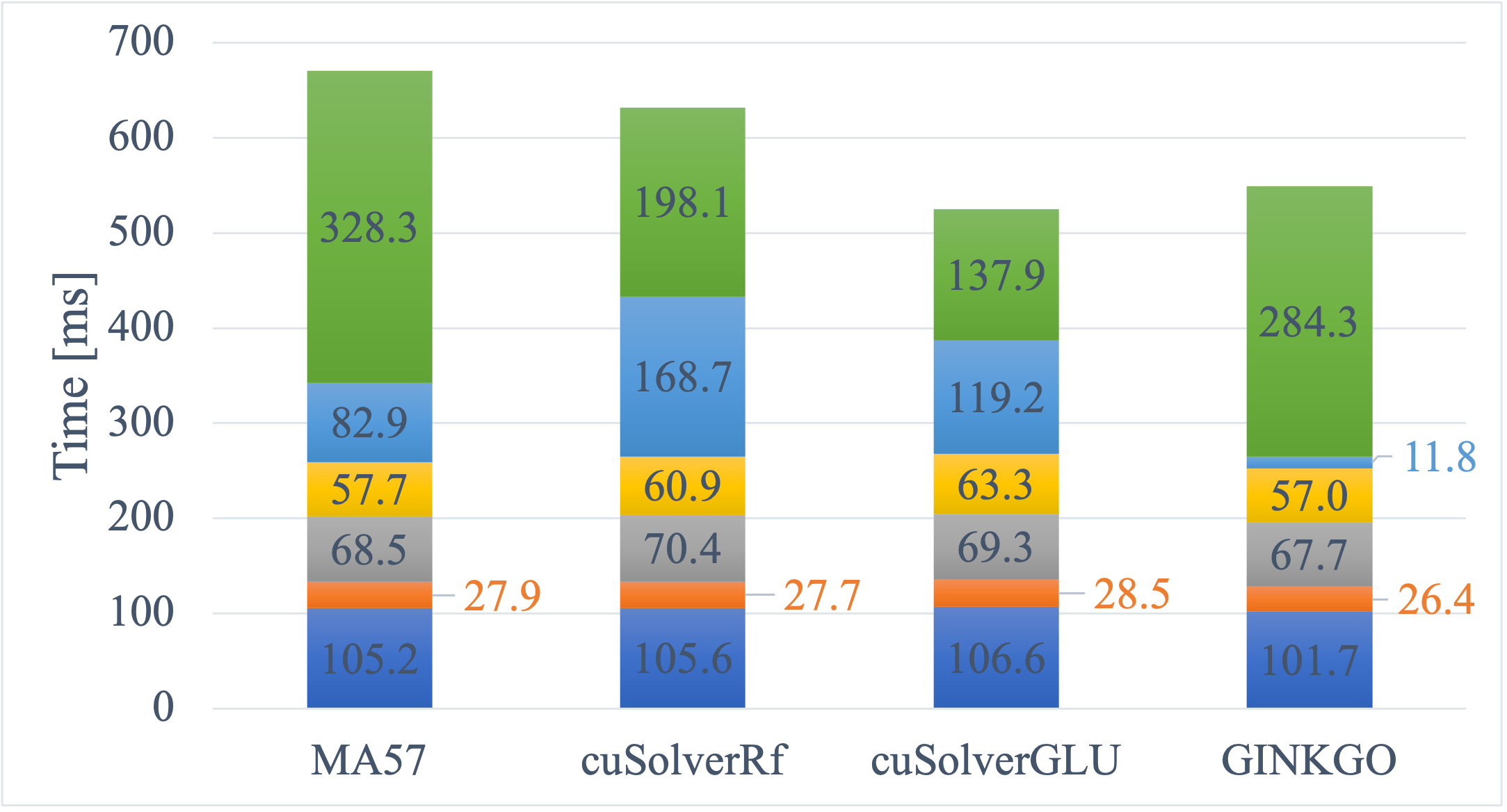}}
    \caption{Eastern U.S. grid}
  \end{subfigure}
  \begin{subfigure}{0.8\columnwidth}
    {\includegraphics[width=\columnwidth,trim={0.1cm 0.1cm 0.1cm 0.1cm},clip]{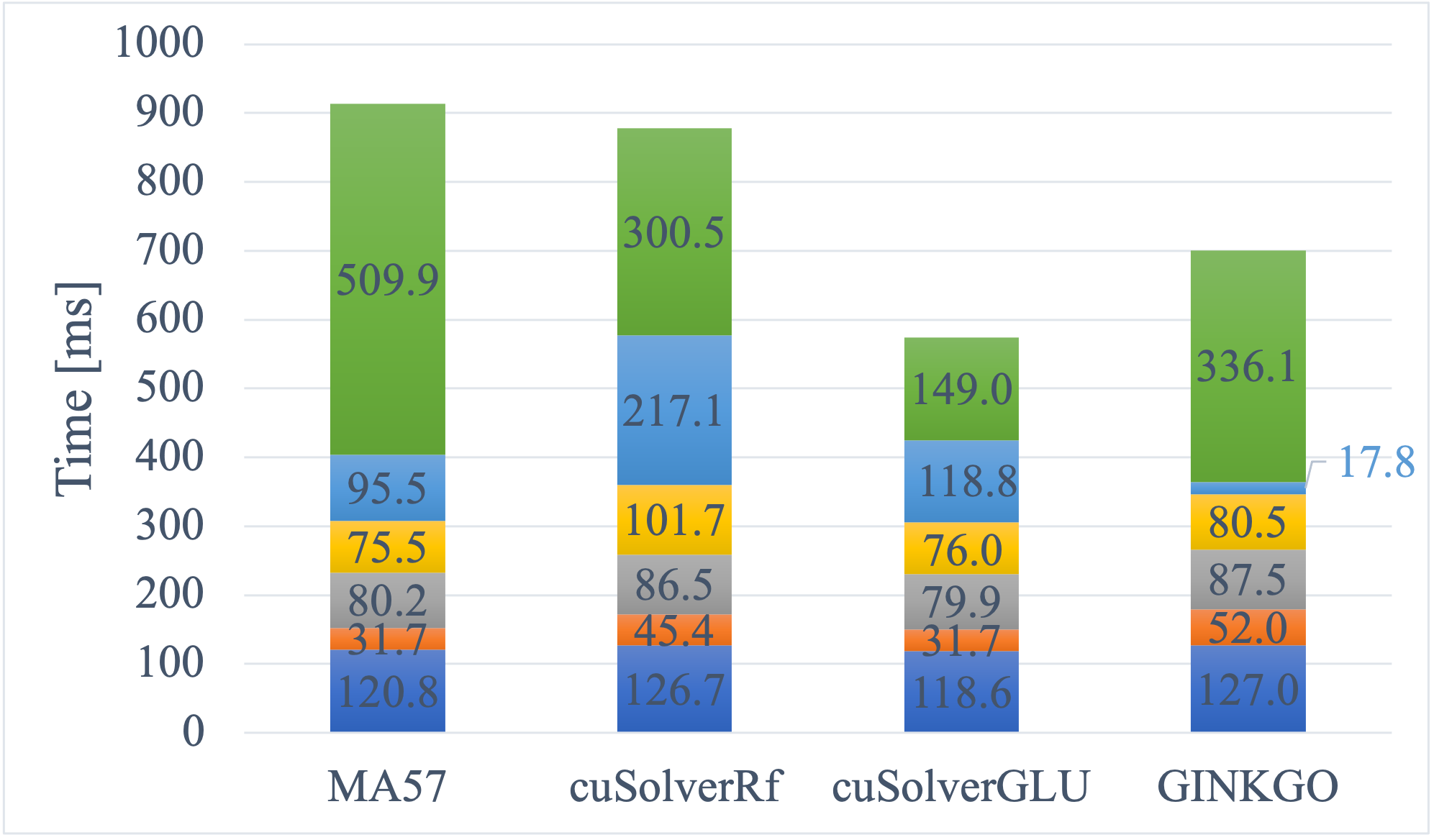}}
    \caption{Eastern and Western U.S. grids}
  \end{subfigure}
  \caption{Comparison of the average computational cost per {optimization solver step} when different linear solvers are used for \ac{acopf} on \ac{olcf} Summit with a breakdown in terms of most expensive operations. The cost of the first step, which is executed on \ac{cpu}, is accounted for in the averages.
  }
  \label{fig:ma57_vs_cusolver_vs_ginkgo}
\end{figure}

\begin{figure}[htb]
\centering
  {\includegraphics[width=\columnwidth,trim={0cm 0cm 0cm 0cm},clip]{./profile_legend}} \\
  \begin{subfigure}{0.45\columnwidth}
    {\includegraphics[width=\columnwidth,trim={0.1cm 0.1cm 0.1cm 0.1cm},clip]{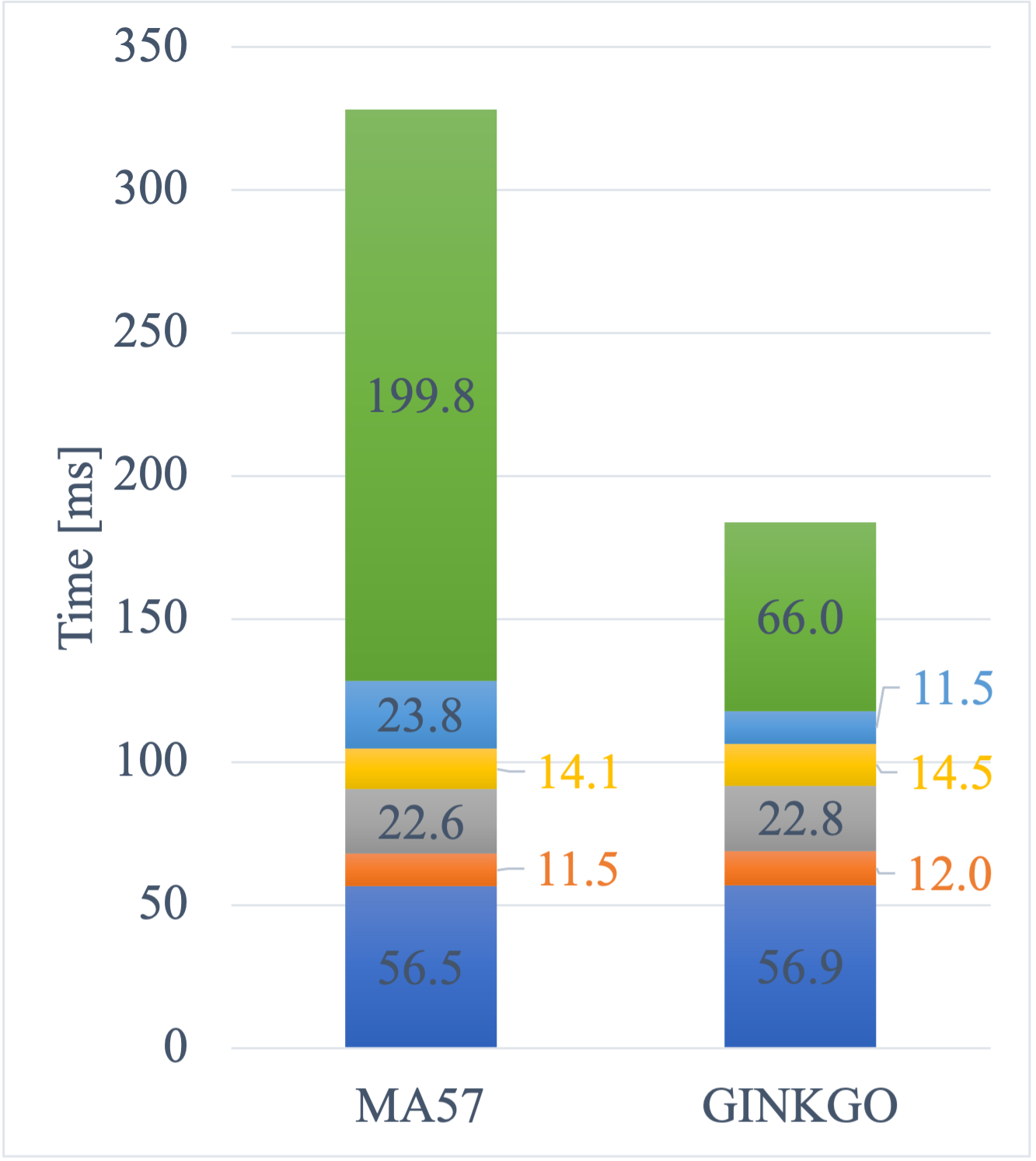}}
    \caption{Northeast U.S. grid}
  \end{subfigure}
  \begin{subfigure}{0.45\columnwidth}
    {\includegraphics[width=\columnwidth,trim={0.1cm 0.1cm 0.1cm 0.1cm},clip]{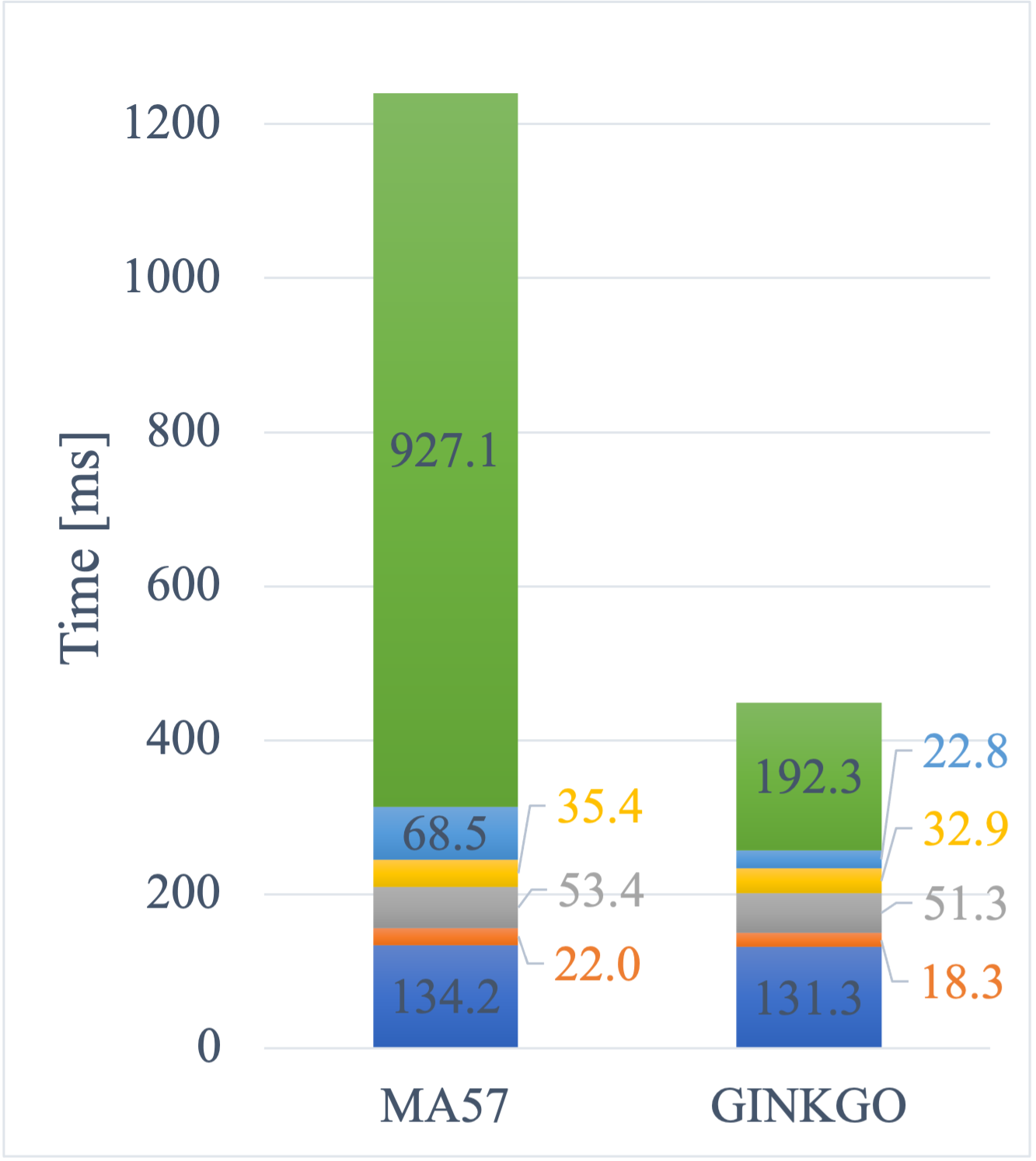}}
    \caption{Eastern U.S. grid}
  \end{subfigure}
  \begin{subfigure}{0.45\columnwidth}
    {\includegraphics[width=\columnwidth,trim={0.1cm 0.1cm 0.1cm 0.1cm},clip]{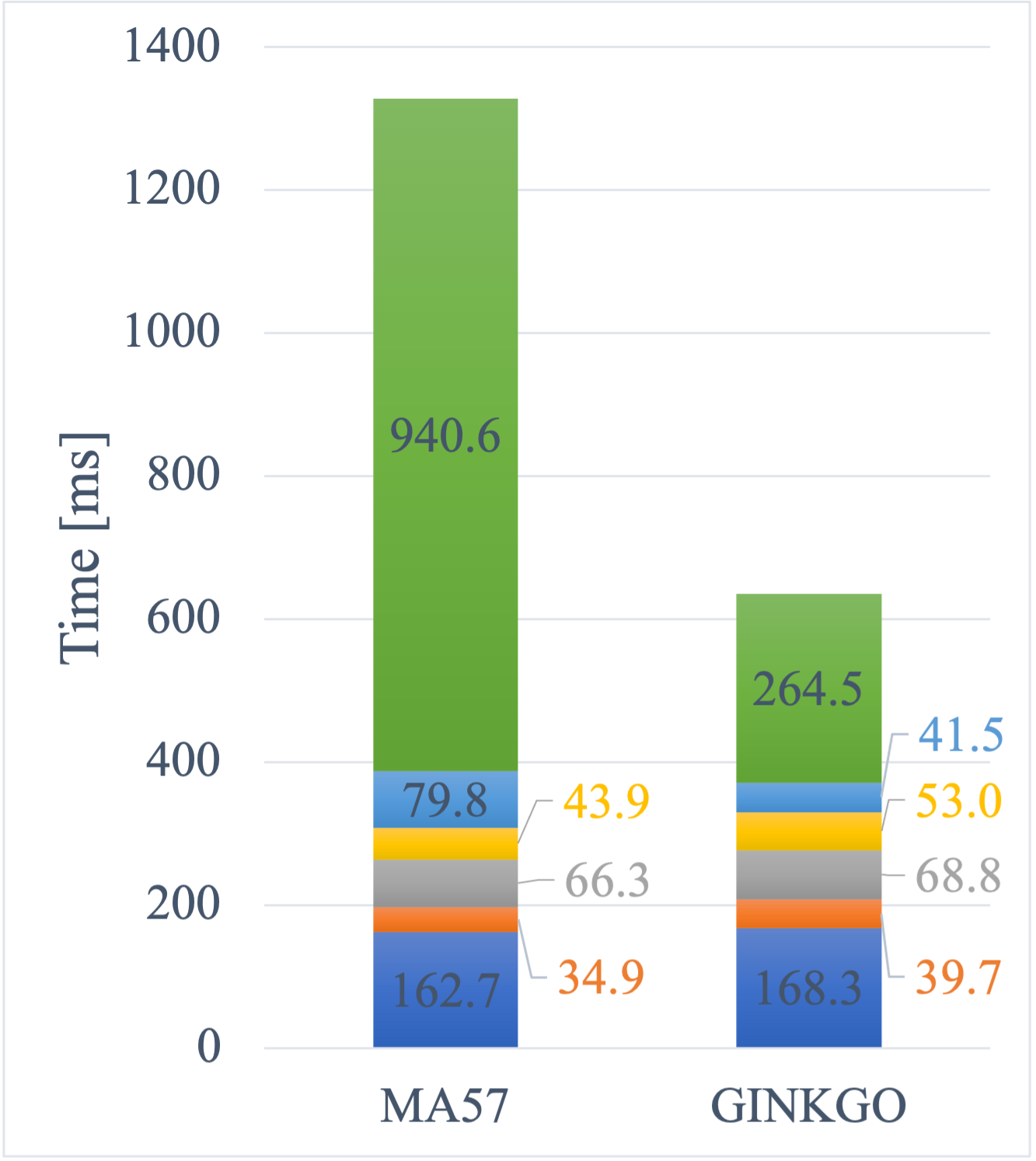}}
    \caption{Eastern and Western U.S. grids}
  \end{subfigure}
  \caption{Comparison of the average computational cost per {optimization solver step} when different linear solvers are used for \ac{acopf} on \ac{olcf} Crusher with a breakdown in terms of most expensive operations. The cost of the first step, which is executed on \ac{cpu}, is accounted for in the averages.
  }
  \label{fig:ma57_vs_ginkgo}
\end{figure}

Figures \ref{fig:ma57_vs_cusolver_vs_ginkgo} and \ref{fig:ma57_vs_ginkgo} provide additional insight by breaking down the average runtime of an optimization solver (in this case \hiop) step into its components. The averages were obtained by normalizing the total computational time for each component by the number of optimization steps to allow comparison between the approaches, and these averages account for the cost of the first factorization performed on \ac{cpu}.

We first focus on the evaluation of the benchmark results obtained on the \ac{olcf} Summit system equipped with \nvidia V100 \ac{gpu}.
The results reported in Table \ref{tab:ma57_vs_cusolver_vs_ginkgo} reveal that the GPU solvers outperform the MA57 baseline implementation for all test cases.  \verb|cuSolverRf| is $10-30\%$ slower overall than \verb|cuSolverGLU|, though it requires fewer iterations overall to converge. This implies that the iterative refinement improves the solution quality and the rate of convergence for an additional computational cost for the cases evaluated.  In terms of overall \ac{acopf} compute time, using \verb|cuSolverGLU| and \gko on V100 \ac{gpu} leads to  $1.3-1.4\times$ and $1.05-1.3 \times$ faster solution, respectively, compared to the \ac{cpu} baseline with MA57. 

The runtime breakdown in Figure \ref{fig:ma57_vs_cusolver_vs_ginkgo} reveals that the performance advantage of the \verb|cuSolverGLU| is primarily driven by a faster factorization (which is for the combined Eastern and Western U.S. grid about 3.4$\times$ faster than MA57).  The faster factorization compensates for the slower triangular solves: The \verb|cuSolverGLU| triangular solves are about 40\% slower than the \ac{cpu} counterpart. For the \gko GPU solver, the story is different: though still faster than the MA57 code, the speedup achieved with the \gko factorization is smaller. On the other hand, the triangular solve in \gko is faster, mainly because it does not call iterative refinement.
We note that it is impossible to combine the \verb|cuSolverGLU| factorization with the \gko sparse triangular solves as the \verb|cuSolverGLU| does not provide access to the triangular factors.

We now turn to the performance results on the \ac{olcf} Crusher system featuring AMD MI250 \acp{gpu}. The results in Table \ref{tab:ma57_vs_ginkgo} reveal that \ac{acopf} is overall $1.8-2.4 \times$ faster when using the \gko linear solver on the AMD MI250X \ac{gpu} than MA57 on the AMD EPYC 7A53 \ac{cpu}. The runtime breakdown in Figure \ref{fig:ma57_vs_ginkgo} reveals that the performance superiority comes from both a faster factorization ($3-4.8 \times$ speedup) and faster triangular solves ($1.9-3 \times$ speedup). Comparing \gko's linear solver (triangular solve and factorization) performance on the MI250X \ac{gpu} and the \nvidia V100 \ac{gpu}, we notice that executing \gko on the newer MI250 \ac{gpu} is 20\% -- 40\% faster than on the \nvidia V100 \ac{gpu}. MA57 runs slower on the AMD EPYC 7A53 \ac{cpu} than on the IBM Power 9 \ac{cpu}.


{ For completeness, we finally discuss the memory requirements for the sparse linear solvers on \ac{gpu}. The \ac{gpu} high bandwidth memory required by the linear solvers for three grids evaluated is less than 3 GB (well below the 16 GB available on NVIDIA V100 and the 64 GB available on a graphics compute die of AMD MI250X). More specifically, on NVIDIA V100, the solver based on \verb|cuSolverGLU| requires 2383-3065 MiB, the solver based on \verb|cuSolverRf| requires 673-1125 MiB, and \textsc{Ginkgo} requires 725-1109 MiB.}

\begin{table}[htb]
    \centering
    \footnotesize
    \caption{Total runtimes for \ac{acopf} when using different linear solvers on \ac{olcf} Summit. The number of steps is the total number of optimization solver steps to the converged solution.}
    \scalebox{0.95}{
    \begin{tabular}{p{0.27\columnwidth}|p{0.08\columnwidth}|p{0.15\columnwidth}|p{0.18\columnwidth}|p{0.09\columnwidth}}
    \hline
    \multicolumn{5}{c}{\textbf{Northeast U.S. grid}} \\
    \hline
     Linear solver used & MA57 & cuSolverRf & cuSolverGLU & \gko \\ \hline
     Total time (s)  & 152 & 127 & 116 & 114 \\ 
     Speedup vs MA57 & - & 1.2 & 1.3 & 1.3 \\ \hline
     Number of steps & 529 & 425 & 547 & 527 \\ \hline
    \multicolumn{5}{c}{\textbf{Eastern U.S. grid}} \\
    \hline
     Linear solver used & MA57 & cuSolverRf & cuSolverGLU & \gko \\ \hline
     Total time (s) & 196 & 187 & 147 & 153 \\ 
     Speedup vs MA57 & - & 1.05 & 1.3 & 1.3 \\ \hline
     Number of steps & 263 & 262 & 263 & 263 \\ \hline
    \multicolumn{5}{c}{\textbf{Combined eastern and western U.S. grids}} \\
    \hline
     Linear solver used & MA57 & cuSolverRf & cuSolverGLU & \gko \\ \hline
     Total time (s) & 793 & 674 & 575 & 766 \\ 
     Speedup vs MA57 & - & 1.2 & 1.4 & 1.04 \\ \hline
     Number of steps & 852 & 735 & 747 & 1038 \\ \hline
    \end{tabular}}
    \label{tab:ma57_vs_cusolver_vs_ginkgo}
\end{table}

\begin{table}[htb]
    \centering
    \footnotesize
    \caption{Total runtimes for \ac{acopf} when using MA57 and \gko linear solvers on \ac{olcf} Crusher. The number of steps is the total number of optimization solver steps to the converged solution.}
    \scalebox{0.95}{
    \begin{tabular}{p{0.35\columnwidth}|p{0.2\columnwidth}|p{0.2\columnwidth}}
    \hline
    \multicolumn{3}{c}{\textbf{Northeast U.S. grid}} \\
    \hline
     Linear solver used & MA57 & \gko \\ \hline
     Total time (s) & 149 & 81 \\ 
     Speedup vs MA57 & - & 1.8 \\ \hline
     Number of steps & 455 & 446 \\ \hline
    \multicolumn{3}{c}{\textbf{Eastern U.S. grid}} \\
    \hline
     Linear solver used & MA57 & \gko \\ \hline
     Total time (s) & 293 &  122 \\ 
     Speedup vs MA57 & - & 2.4 \\ \hline
     Number of steps & 234 & 263 \\ \hline
    \multicolumn{3}{c}{\textbf{Combined eastern and western U.S. grids}} \\
    \hline
     Linear solver used & MA57 & \gko \\ \hline
     Total time (s) & 927 &  450 \\ 
     Speedup vs MA57 & - & 2.1 \\ \hline
     Number of steps & 693 & 700 \\ \hline
    \end{tabular}}
    \label{tab:ma57_vs_ginkgo}
\end{table}

\section{Conclusions and Next Steps}
\label{sec:conclusion}

This paper follows up on the analysis from~\cite{swirydowicz2022linear} and presents different strategies to overcome the challenge of fast GPU-resident sparse direct solvers. Unlike in~\cite{swirydowicz2022linear}, we test linear solvers within full optimal power flow applications at interconnection scale, not on standalone test matrices. Our performance analysis reveals that the linear solver dominates the overall performance.

On both the IBM/NVIDIA-based Summit supercomputer and the AMD-based Crusher cluster, the \ac{gpu}-based sparse linear solvers are faster than the \ac{cpu}-based solvers. 
Furthermore, \gko's GPU-resident sparse direct solver functionality brings platform portability to the \ac{acopf} simulation and is, for all considered hardware configurations, faster than the \ac{cpu}-based counterparts. 
For our test cases, we observed $1.9-2.3\times$ linear solver speedup when comparing the best \ac{gpu} to the best \ac{cpu} result (see Figures \ref{fig:ma57_vs_cusolver_vs_ginkgo} and \ref{fig:ma57_vs_ginkgo}). This projects to $1.6$--$1.9\times$ overall speedup in \ac{acopf} analysis when comparing the best \ac{gpu} to the best \ac{cpu} time (see Tables \ref{tab:ma57_vs_cusolver_vs_ginkgo} and \ref{tab:ma57_vs_ginkgo}). 
These and the results in \cite{abhyankar2021acopf} suggest it is feasible to develop methods for efficient \ac{acopf} on \acp{gpu} and obtain significant performance gains.

We note that the refactorization approaches we employ in this work {reuse the same pivot sequence over a large number of linear systems (\ref{eq:kktlinear})}, and hence come with the risk of deterioration in solution quality. 
On the other hand, the linear solver error only needs to be small compared to the allowed error in optimization solver iteration. We find that with \hiop's default tolerance setting of $10^{-8}$, refactorization approach without iterative refinement as implemented in \gko will perform well in most cases. The exception is the \ac{acopf} analysis for the Eastern and Western U.S. grid, where analysis using \gko requires nearly 200 more optimization solver steps than the reference analysis using MA57 linear solver on \ac{cpu} (see Table \ref{tab:ma57_vs_cusolver_vs_ginkgo}).

We observed that using high precision iterative refinement with \verb|cuSolverRf| leads to fewer optimization solver steps to the solution as expected (Table \ref{tab:ma57_vs_cusolver_vs_ginkgo}). However, iterative refinement can add significant computational overhead, as shown in Figure \ref{fig:ma57_vs_cusolver_vs_ginkgo}, as it calls triangular solver repeatedly. Finding the right balance between reducing the number of optimization solver steps and the cost of each step in order to reduce the overall computational cost is a nontrivial problem. We will investigate this further in the follow on research.

We find that numerical (re)factorization in \verb|cuSolverGLU| is superior to those in \verb|cuSolverRf| and \gko. However, \verb|cuSolverGLU| does not provide access to the $L$ and $U$ matrix factors, so its factorization function cannot be combined with other solvers. At the same time, its triangular solver function performs poorly compared to MA57 reference, most likely due to built-in iterative refinement, which the user cannot configure or disable. As a next step, we will work towards improving the performance of \gko refactorization functions.

Future work will also include algorithmic and technical strategies to speed up triangular solves by using an iterative approximation as proposed in~\cite{anzt2015iterative}. Furthermore, we will investigate the hybrid strategy developed in \cite{regev2022kkt}, which applies both iterative and direct solvers with very promising performance results.
Hybrid strategies will likely become more attractive in the future as the hardware manufacturers plan to develop processing units with \ac{cpu} and \ac{gpu} cores on the same die.

Finally, the full advantage of deploying \ac{acopf} on heterogeneous hardware will become apparent when the entire analysis is ported to \ac{gpu}, including the model evaluation and optimization solver. We demonstrated this in \cite{abhyankar2021acopf} where the entire compressed \ac{acopf} model and dense linear solver were run on the GPU. Sparse \ac{gpu} modules are currently being developed in both, \hiop and \exago libraries and, when completed, we will use them to deploy the entire analysis on \ac{gpu}. 

\section*{Acknowledgments}
\noindent
This research has been supported in part by UT-Battelle, LLC, and used resources of the Oak Ridge Leadership Computing Facility under contract DE-AC05-00OR22725 with the U.S. Department of Energy (DOE). This research was also supported by the Exascale Computing Project (17-SC-20-SC), a collaborative effort of the DOE Office of Science and the National Nuclear Security Administration. 

The authors thank Cosmin Petra and Nai-Yuan Chiang of Lawrence Livermore National Laboratory for their guidance when using \hiop optimization solver.
Warm thanks also go to Phil Roth of Oak Ridge National Laboratory and Christopher Oehmen of Pacific Northwest National Laboratory for their support of this work. 

\section*{Notice of Copyright}
\noindent
This manuscript has been authored in part by UT-Battelle, LLC, under contract DE-AC05-00OR22725 with the US Department of Energy (DOE). The US government retains and the publisher, by accepting the article for publication, acknowledges that the US government retains a nonexclusive, paid-up, irrevocable, worldwide license to publish or reproduce the published form of this manuscript, or allow others to do so, for US government purposes. DOE will provide public access to these results of federally sponsored research in accordance with the DOE Public Access Plan (http://energy.gov/downloads/doe-public-access-plan).

\appendix


\bibliographystyle{elsarticle-num} 
\bibliography{refs}

\begin{thebibliography}{10}
\expandafter\ifx\csname url\endcsname\relax
  \def\url#1{\texttt{#1}}\fi
\expandafter\ifx\csname urlprefix\endcsname\relax\def\urlprefix{URL }\fi
\expandafter\ifx\csname href\endcsname\relax
  \def\href#1#2{#2} \def\path#1{#1}\fi

\bibitem{ONeill2012}
R.~P. O'Neill, A.~Castillo, M.~B. Cain,
  \href{http://www.ferc.gov/industries/electric/indus-act/market-planning/opf-papers/acopf-2-iv-linearization.pdf}{{The
  IV formulation and linear approximations of the AC optimal power flow problem
  (OPF Paper 2)}}, FERC Staff Technical Paper~(December) (2012) 1--18.
\newline\urlprefix\url{http://www.ferc.gov/industries/electric/indus-act/market-planning/opf-papers/acopf-2-iv-linearization.pdf}

\bibitem{Frank2012}
S.~Frank, S.~Rebennack, et~al., {A Primer on Optimal Power Flow: Theory,
  Formulation, and Practical Examples}, Tech. Rep.~14, Colorado School of Mines
  (2012).

\bibitem{swirydowicz2022linear}
K.~{\'S}wirydowicz, E.~Darve, W.~Jones, J.~Maack, S.~Regev, M.~A. Saunders,
  S.~J. Thomas, S.~Pele{\v{s}}, Linear solvers for power grid optimization
  problems: a review of {GPU}-accelerated linear solvers, Parallel Computing
  111 (2022) 102870.

\bibitem{dinkelbach2021factorisation}
J.~Dinkelbach, L.~Schumacher, L.~Razik, A.~Benigni, A.~Monti, Factorisation
  path based refactorisation for high-performance {LU} decomposition in
  real-time power system simulation, Energies 14~(23) (2021) 7989.

\bibitem{razik2019comparative}
L.~Razik, L.~Schumacher, A.~Monti, A.~Guironnet, G.~Bureau, A comparative
  analysis of {LU} decomposition methods for power system simulations, in: 2019
  IEEE Milan PowerTech, IEEE, 2019, pp. 1--6.

\bibitem{dorto2021comparing}
M.~D’orto, S.~Sj{\"o}blom, L.~S. Chien, L.~Axner, J.~Gong, Comparing
  different approaches for solving large scale power-flow problems with the
  {N}ewton-{R}aphson method, IEEE Access 9 (2021) 56604--56615.

\bibitem{Rakai2014}
L.~Rakai, W.~Rosehart, {GPU}-accelerated solutions to optimal power flow
  problems, in: 2014 47th Hawaii International Conference on System Sciences,
  IEEE, 2014, pp. 2511--2516.

\bibitem{abhyankar2021acopf}
S.~Abhyankar, S.~Peles, R.~Rutherford, A.~Mancinelli, Evaluation of {AC}
  optimal power flow on graphical processing units, in: 2021 IEEE Power \&
  Energy Society General Meeting (PESGM), 2021, pp. 01--05.
\newblock \href {https://doi.org/10.1109/PESGM46819.2021.9638131}
  {\path{doi:10.1109/PESGM46819.2021.9638131}}.

\bibitem{su2020full}
X.~Su, C.~He, T.~Liu, L.~Wu, Full parallel power flow solution: A
  {GPU-CPU}-based vectorization parallelization and sparse techniques for
  {N}ewton--{R}aphson implementation, IEEE Transactions on Smart Grid 11~(3)
  (2020) 1833--1844.

\bibitem{nvidia2021cusolver}
NVIDIA, {cuSOLVER} {L}ibrary, release 11.4 (2021).

\bibitem{anzt2022ginkgo}
H.~Anzt, T.~Cojean, G.~Flegar, F.~Göbel, T.~Grützmacher, P.~Nayak,
  T.~Ribizel, Y.~M. Tsai, E.~S. Quintana-Ortí, {Ginkgo: A Modern Linear
  Operator Algebra Framework for High Performance Computing}, ACM Transactions
  on Mathematical Software 48~(1) (2022) 2:1--2:33.
\newblock \href {https://doi.org/10.1145/3480935} {\path{doi:10.1145/3480935}}.

\bibitem{Zimmerman2011}
R.~D. Zimmerman, C.~E. Murillo-S{\'{a}}nchez, R.~J. Thomas, {MATPOWER:
  Steady-state operations, planning, and analysis tools for power systems
  research and education}, IEEE Transactions on Power Systems 26~(1) (2011)
  12--19.
\newblock \href {https://doi.org/10.1109/TPWRS.2010.2051168}
  {\path{doi:10.1109/TPWRS.2010.2051168}}.

\bibitem{wachter2006implementation}
A.~W{\"a}chter, L.~T. Biegler, On the implementation of an interior-point
  filter line-search algorithm for large-scale nonlinear programming,
  Mathematical programming 106~(1) (2006) 25--57.

\bibitem{birchfield2017tamu-cases}
A.~B. {Birchfield}, T.~{Xu}, K.~M. {Gegner}, K.~S. {Shetye}, T.~J. {Overbye},
  Grid structural characteristics as validation criteria for synthetic
  networks, IEEE Transactions on Power Systems 32~(4) (2017) 3258--3265.
\newblock \href {https://doi.org/10.1109/TPWRS.2016.2616385}
  {\path{doi:10.1109/TPWRS.2016.2616385}}.

\bibitem{duff2004ma57}
I.~S. Duff, {MA57}---a code for the solution of sparse symmetric definite and
  indefinite systems, ACM Transactions on Mathematical Software (TOMS) 30~(2)
  (2004) 118--144.

\bibitem{ExaGo}
S.~Abhyankar, S.~Peles, A.~Mancinelli, R.~Rutherford, B.~Palmer,
  \href{https://https://gitlab.pnnl.gov/exasgd/frameworks/exago}{Exascale grid
  optimization toolkit} (2020).
\newline\urlprefix\url{https://https://gitlab.pnnl.gov/exasgd/frameworks/exago}

\bibitem{hiop_techrep}
C.~G. Petra, N.~Chiang, J.~Wang, {HiOp} -- {U}ser {G}uide, Tech. Rep.
  LLNL-SM-743591, Center for Applied Scientific Computing, Lawrence Livermore
  National Laboratory (2018).

\bibitem{chiang2016inertia}
N.-Y. Chiang, V.~M. Zavala, An inertia-free filter line-search algorithm for
  large-scale nonlinear programming, Computational Optimization and
  Applications 64~(2) (2016) 327--354.

\bibitem{li2003superlu_dist}
X.~S. Li, J.~W. Demmel, Superlu\_dist: A scalable distributed-memory sparse
  direct solver for unsymmetric linear systems, ACM Transactions on
  Mathematical Software (TOMS) 29~(2) (2003) 110--140.

\bibitem{ghysels2016efficient}
P.~Ghysels, X.~S. Li, F.-H. Rouet, S.~Williams, A.~Napov, An efficient
  multicore implementation of a novel {HSS}-structured multifrontal solver
  using randomized sampling, SIAM Journal on Scientific Computing 38~(5) (2016)
  S358--S384.

\bibitem{hogg2016sparse}
J.~D. Hogg, E.~Ovtchinnikov, J.~A. Scott, A sparse symmetric indefinite direct
  solver for {GPU} architectures, ACM Transactions on Mathematical Software
  (TOMS) 42~(1) (2016) 1--25.

\bibitem{li05}
X.~S. Li, An overview of {SuperLU}: Algorithms, implementation, and user
  interface, ACM Trans. Math. Softw. 31~(3) (2005) 302--325.

\bibitem{Duff2020}
I.~Duff, J.~Hogg, F.~Lopez, A new sparse {$\textit{LDL}^\textit{T}$} solver
  using a posteriori threshold pivoting, SIAM Journal on Scientific Computing
  42~(2) (2020) C23--C42.
\newblock \href {https://doi.org/10.1137/18M1225963}
  {\path{doi:10.1137/18M1225963}}.

\bibitem{henon2002pastix}
P.~H{\'e}non, P.~Ramet, J.~Roman, {PaStiX}: a high-performance parallel direct
  solver for sparse symmetric positive definite systems, Parallel Computing
  28~(2) (2002) 301--321.

\bibitem{davis2010algorithm}
T.~A. Davis, E.~Palamadai~Natarajan, Algorithm 907: {KLU}, a direct sparse
  solver for circuit simulation problems, ACM Transactions on Mathematical
  Software (TOMS) 37~(3) (2010) 1--17.

\bibitem{chen2013nicslu}
X.~Chen, Y.~Wang, H.~Yang, {NICSLU:} an adaptive sparse matrix solver for
  parallel circuit simulation, IEEE transactions on computer-aided design of
  integrated circuits and systems 32~(2) (2013) 261--274.

\bibitem{he2015gpu}
K.~He, S.~X.-D. Tan, H.~Wang, G.~Shi, {GPU}-accelerated parallel sparse {LU}
  factorization method for fast circuit analysis, IEEE Transactions on Very
  Large Scale Integration (VLSI) Systems 24~(3) (2015) 1140--1150.

\bibitem{duff2007static}
I.~S. Duff, S.~Pralet, Towards stable mixed pivoting strategies for the
  sequential and parallel solution of sparse symmetric indefinite systems, SIAM
  Journal on Matrix Analysis and Applications 29~(3) (2007) 1007--1024.
\newblock \href {https://doi.org/10.1137/050629598}
  {\path{doi:10.1137/050629598}}.

\bibitem{AMD_reordering}
P.~R. Amestoy, T.~A. Davis, I.~S. Duff, Algorithm 837: {AMD}, {A}n
  {A}pproximate {M}inimum {D}egree {O}rdering {A}lgorithm, ACM Trans. Math.
  Softw. 30~(3) (2004) 381–388.
\newblock \href {https://doi.org/10.1145/1024074.1024081}
  {\path{doi:10.1145/1024074.1024081}}.

\bibitem{wilkinson1965rounding}
J.~Wilkinson, Rounding errors in algebraic processes (1965).

\bibitem{saad2003iterative}
Y.~Saad, Iterative methods for sparse linear systems, SIAM, 2003.

\bibitem{saad1993flexible}
Y.~Saad, A flexible inner-outer preconditioned {GMRES} algorithm, SIAM Journal
  on Scientific Computing 14~(2) (1993) 461--469.

\bibitem{arioli2007note}
M.~Arioli, I.~S. Duff, S.~Gratton, S.~Pralet, A note on {GMRES} preconditioned
  by a perturbed $\textit{LDL}^\textit{T}$ decomposition with static pivoting,
  SIAM Journal on Scientific Computing 29~(5) (2007) 2024--2044.

\bibitem{carson2020three}
E.~Carson, N.~J. Higham, S.~Pranesh, Three-precision {GMRES}-based iterative
  refinement for least squares problems, SIAM Journal on Scientific Computing
  42~(6) (2020) A4063--A4083.

\bibitem{mc64}
I.~S. Duff, J.~Koster, On algorithms for permuting large entries to the
  diagonal of a sparse matrix, SIAM Journal on Matrix Analysis and Applications
  22~(4) (2001) 973--996.
\newblock \href {https://doi.org/10.1137/S0895479899358443}
  {\path{doi:10.1137/S0895479899358443}}.

\bibitem{ginkgo-release}
{Ginkgo Project},
  \href{https://github.com/ginkgo-project/ginkgo/releases/tag/v1.6.0}{Ginkgo
  release 1.6.0} (2023).
\newline\urlprefix\url{https://github.com/ginkgo-project/ginkgo/releases/tag/v1.6.0}

\bibitem{rose_algorithmic_1978}
D.~J. Rose, R.~E. Tarjan, Algorithmic {Aspects} of {Vertex} {Elimination} on
  {Directed} {Graphs}, SIAM Journal on Applied Mathematics 34~(1) (1978)
  176--197.
\newblock \href {https://doi.org/10.1137/0134014} {\path{doi:10.1137/0134014}}.

\bibitem{liu2016syncfree}
W.~Liu, A.~Li, J.~Hogg, I.~S. Duff, B.~Vinter, A synchronization-free algorithm
  for parallel sparse triangular solves, in: P.-F. Dutot, D.~Trystram (Eds.),
  Euro-Par 2016: Parallel Processing, Springer International Publishing, Cham,
  2016, pp. 617--630.

\bibitem{gkoportability}
T.~Cojean, Y.-H.~M. Tsai, H.~Anzt, Ginkgo - a math library designed for
  platform portability, Parallel Computing 111 (2022) 102902.
\newblock \href {https://doi.org/10.1016/j.parco.2022.102902}
  {\path{doi:10.1016/j.parco.2022.102902}}.

\bibitem{anzt2015iterative}
H.~Anzt, E.~Chow, J.~Dongarra, Iterative sparse triangular solves for
  preconditioning, in: European conference on parallel processing, Springer,
  2015, pp. 650--661.

\bibitem{regev2022kkt}
S.~Regev, N.-Y. Chiang, E.~Darve, C.~G. Petra, M.~A. Saunders,
  K.~\'{S}wirydowicz, S.~Pele\v{s}, {HyKKT:} a hybrid direct-iterative method
  for solving {KKT} linear systems, Optimization Methods and Software (2022)
  1--24\href {https://doi.org/10.1080/10556788.2022.2124990}
  {\path{doi:10.1080/10556788.2022.2124990}}.

\end{thebibliography}





\end{document}